\documentclass[aps,prc,showpacs,floatfix,footinbib,twocolumn]{revtex4}
\usepackage{amsmath}
\usepackage[dvips]{graphics,graphicx}
\usepackage{graphicx}
\usepackage{dcolumn}
\usepackage{bm}

\begin{document}


\title{Interacting hadron resonance gas model in K-matrix formalism}

\author{Ashutosh Dash}
\email{ashutosh.dash@niser.ac.in}

\author{Subhasis Samanta}
\email{subhasis.samant@gmail.com}

\author{Bedangadas Mohanty}
\email{bedanga@niser.ac.in}

\affiliation{School of Physical Sciences, National Institute of Science 
Education and Research, HBNI, Jatni - 752050, India}

\begin{abstract}
An extension of Hadron Resonance Gas (HRG) model is constructed to
include interactions using relativistic virial expansion of partition
function. The non-interacting part of the expansion contains all the
stable baryons and mesons and the interacting part contains all the
higher mass resonances which decay into two stable hadrons. The virial coefficients are related to the phase shifts which are calculated using
K-matrix formalism in the present work. 
We have calculated various thermodynamics quantities like pressure,
energy density, and entropy density of the 
system. A 
comparison of thermodynamic quantities with non interacting HRG model, calculated using the same number of 
hadrons, shows that the results of above formalism are larger.
A good agreement between equation of state calculated in K-matrix formalism and lattice QCD simulations
is observed. Specifically the lattice QCD calculated interaction measure is well described in
our formalism. 
We have also calculated second order fluctuations and correlations of conserved charges 
in K-matrix formalism. We observe a good agreement of second order fluctuations and baryon-strangeness 
correlation with lattice data below the cross-over temperature. 
\end{abstract}

\maketitle

\section{Introduction}

Relativistic heavy ion collisions have contributed immensely to our understanding of strongly interacting matter
at finite temperature ($T$) and baryon chemical potential (${\mu}_B$).
Lattice quantum chromodynamics (LQCD) ~\cite{Aoki:2006we,Borsanyi:2011sw, Gupta:2011wh, Bazavov:2012jq,Bellwied:2013cta, 
Borsanyi:2013bia, Bazavov:2014pvz, Bellwied:2015lba}
calculation provides a first principle approach to study strongly interacting matter
at zero baryon chemical potential ($\mu_B$) and finite temperature ($T$)
which indicates a smooth cross over transition ~\cite{Aoki:2006we} from 
hadronic to a quark-gluon plasma (QGP) phase \cite{Gupta:2011wh}.
On the other hand, at high baryon chemical potential the nuclear matter is expected to have a
first-order phase transition ~\cite{Asakawa:1989bq} which ends
at a critical point. Several experimental program have been devoted to study strongly interacting matter in a wide range
of temperature and baryon chemical potential. At present, the
properties of matter at  high temperature and small baryon chemical potential are being investigated using ultra relativistic 
heavy ion collisions at the Large Hadron Collider (LHC), CERN
and Relativistic Heavy Ion Collider (RHIC), Brookhaven National Laboratory (BNL).
The Beam Energy Scan (BES) program of RHIC \cite{Abelev:2009bw} is currently
investigating the matter at large baryon chemical potential and the location of the critical
point \cite{Adamczyk:2013dal}.
The HADES experiment at GSI, Darmstadt is also investigating a medium with very large baryon chemical potential
\cite{Agakishiev:2015bwu}.
In future, the Compressed Baryonic Matter (CBM) experiment \cite{Ablyazimov:2017guv}
at the Facility for Antiproton and
Ion Research (FAIR) at GSI and the Nuclotron-based Ion Collider 
fAcility (NICA) \cite{Kekelidze:2017ual} at JINR, 
Dubna will study nuclear matter at large baryon chemical potential. 

Hadron resonance gas  
 ~\cite{Hagedorn:1980kb, Rischke:1991ke,Cleymans:1992jz, BraunMunzinger:1994xr, Cleymans:1996cd, Yen:1997rv,
 BraunMunzinger:1999qy, Cleymans:1999st, BraunMunzinger:2001ip, BraunMunzinger:2003zd, Karsch:2003zq, Tawfik:2004sw,
Becattini:2005xt,  Andronic:2005yp, Andronic:2008gu,Begun:2012rf, Andronic:2012ut,
Tiwari:2011km, Fu:2013gga, Tawfik:2013eua, Garg:2013ata, Bhattacharyya:2013oya,
Bhattacharyya:2015zka,Chatterjee:2013yga,Chatterjee:2014ysa,Chatterjee:2014lfa,Becattini:2012xb,Bugaev:2013sfa,
Petran:2013lja, Vovchenko:2014pka, Kadam:2015xsa, Kadam:2015fza, Albright:2014gva, Albright:2015uua,
Bhattacharyya:2015pra, Kapusta:2016kpq, Begun:2016cva,Adak:2016jtk, Xu:2016skm,Fu:2016baf,
Vovchenko:2015xja, Vovchenko:2015vxa, Vovchenko:2015pya, Broniowski:2015oha, Vovchenko:2015idt,
Redlich:2016dpb, Vovchenko:2016rkn, Alba:2016fku, Samanta:2017kmg, Samanta:2017ohm, Sarkar:2017ijd, 
Bhattacharyya:2017gwt, Chatterjee:2017yhp, Alba:2016hwx, Alba:2017bbr,Samanta:2017yhh}
is a popular model to study the QCD matter formed in heavy-ion
collisions at finite temperature and chemical potential. Varieties of HRG models exist in the literature,
some of the which consider interaction between hadrons and some do
not. The ideal hadron resonance gas (HRG) model is
successful in reproducing the zero chemical potential LQCD data of bulk properties of the QCD matter at
moderate temperatures $T \approx 150$ MeV
~\cite{Borsanyi:2011sw, Bazavov:2012jq, Bazavov:2014pvz,Bellwied:2013cta, Bellwied:2015lba}.
This model is also successful in describing the hadron yields, at chemical freezeout, 
created in central heavy ion collisions from SIS up to RHIC energies
~\cite{BraunMunzinger:1994xr, Cleymans:1996cd, Cleymans:1999st, 
Andronic:2005yp}. The ideal HRG model assumes that microscopic thermal system consist of non-interacting point like
hadrons and resonances, hence the width of the resonances are
ignored. There are several approaches to include interaction in the HRG model. 
One such model is excluded volume HRG (EVHRG) model where van der Waals type repulsive 
interaction ~\cite{Hagedorn:1980kb, Rischke:1991ke,
Cleymans:1992jz, Yen:1997rv, BraunMunzinger:1999qy,Tiwari:2011km,
Begun:2012rf, Andronic:2012ut, Fu:2013gga, Tawfik:2013eua, Bhattacharyya:2013oya,Albright:2014gva,Vovchenko:2014pka,
Albright:2015uua,Alba:2016hwx,Kadam:2015xsa, Kadam:2015fza, 
Kapusta:2016kpq} is introduced by considering the geometrical sizes of the hadrons. 
However the long distance repulsive interactions are ignored in this
model. Another major issue of the model is fixing the radii of various hadrons.
In Refs. \cite{Andronic:2012ut,Bhattacharyya:2013oya} it was shown that the LQCD data of 
different thermodynamic quantities can be described in EVHRG model with the fixed
radius parameter between $0.2 -0.3$ fm. The mass dependent hadronic radius is also considered in EVHRG model to study the hadronic multiplicities
at LHC energy and a reasonable agreement between model and experimental data
is found \cite{Alba:2016hwx}. Repulsive interaction can also be 
introduced via repulsive mean field approach ~\cite{Olive:1980dy,Olive:1982we}.
The van der Waals (VDW) type interaction with both attractive and repulsive parts have been introduced recently
in HRG model ~\cite{Vovchenko:2015xja, Vovchenko:2015vxa, Vovchenko:2015pya, Redlich:2016dpb, 
Vovchenko:2016rkn, Samanta:2017yhh}. Such a model introduces more
parameters and fixing them using existing information has its own drawbacks.
The two van der Waals parameters can be fixed either 
by reproducing the properties of the  nuclear matter at zero temperature  ~\cite{Vovchenko:2015vxa}
or by fitting the LQCD data at zero chemical potential
~\cite{Samanta:2017yhh}. In addition to that, the VDWHRG model does
not even include the interactions of mesons since the number densities 
diverge when the chemical potential becomes comparable to the meson mass \cite{Vovchenko:2015xja}.
Compared to ideal HRG model both EVHRG and VDWHRG describes better
the lattice QCD data in the cross-over region.
As discussed above, in lieu of introducing interactions both the
interacting HRG models bring in additional parameters compared to the ideal HRG 
model. The assumptions involved in fixing the additional parameters in
the interacting HRG models are debatable.  \par

Another approach to include interaction in a system consisting of hadronic gas is the S-matrix 
approach \cite{Dashen:1969ep}. This approach expands the partition function using relativistic virial expansion.
The virial coefficients are related to phase shifts, which needs to calculated either 
theoretically \cite{Weinhold:1997ig,Dobado:1998tv} or obtained from experiments. Type of the interaction depends on the sign of the derivative of
phase shift. A positive sign corresponds to the attractive
interaction and a negative sign corresponds to the repulsive
interaction. For an example, the authors of 
\cite{Venugopalan:1992hy} had found that
$\pi-\pi$ channel has the attractive $\delta_0^0$, $\delta_1^1$ phase 
shifts (phase shift is defined as $\delta_l^I$ where
$l$ is the orbital angular momentum and $I$ is the isospin of the channel) and also the repulsive 
$\delta_0^2$ phase shift.  While
Refs. \cite{Venugopalan:1992hy,Welke:1990za,Friman:2015zua,Huovinen:2016xxq} used
experimental phase shifts in their study. A theoretical way of calculating phase shifts is to use the K-matrix formalism \cite{Badalian:1981xj,Wiranata:2013oaa}. The resonances, contributing to the interaction, appear as a sum of poles in the K-matrix. 
This approach preserves the unitarity of S-matrix and neatly handles multiple resonances unlike the popular Breit-Wigner parametrization of the resonance spectral function. 
We would like to mention here that Refs.
\cite{BraunMunzinger:1994xr,Brown:1991en,Sollfrank:1990qz,Gorenstein:1987zm,Andronic:2005yp} used Breit-Wigner parametrization with an ad hoc profile function \cite{Gorenstein:1987zm}. 
All these issues motivates us to use the K-matrix formalism
consistently to calculate phase shifts in the virial expansion approach. 
The K-matrix formalism has been applied previously to calculate shear viscosity and interaction measure for interacting hadronic gas in \cite{Wiranata:2013oaa}. However, our result on interacting measure 
agrees better with the lattice QCD result on including additional resonances. Further, we calculate susceptibilities of the conserved charges within the K-matrix formalism.

The paper is organized as follows. In the Secs. \ref{Sc: K-matrix} and \ref{Sc: BW} we discuss K-martix
formalism and the Breit-Wigner parameterization of the resonance
spectral function respectively. A comparison between above approaches is given
in Sec. \ref{Sc: Comparison}. Section \ref{Sc: VE} discusses relativistic virial expansion.
We discuss numerical results and comparison of our calculations with
ideal HRG and LQCD in Sec \ref{Sc: Results}.  We conclude our findings in Sec. \ref{Sc: Conclusion}.

\begin{figure*}[t]
\begin{center}
\includegraphics[width=0.41\textwidth]{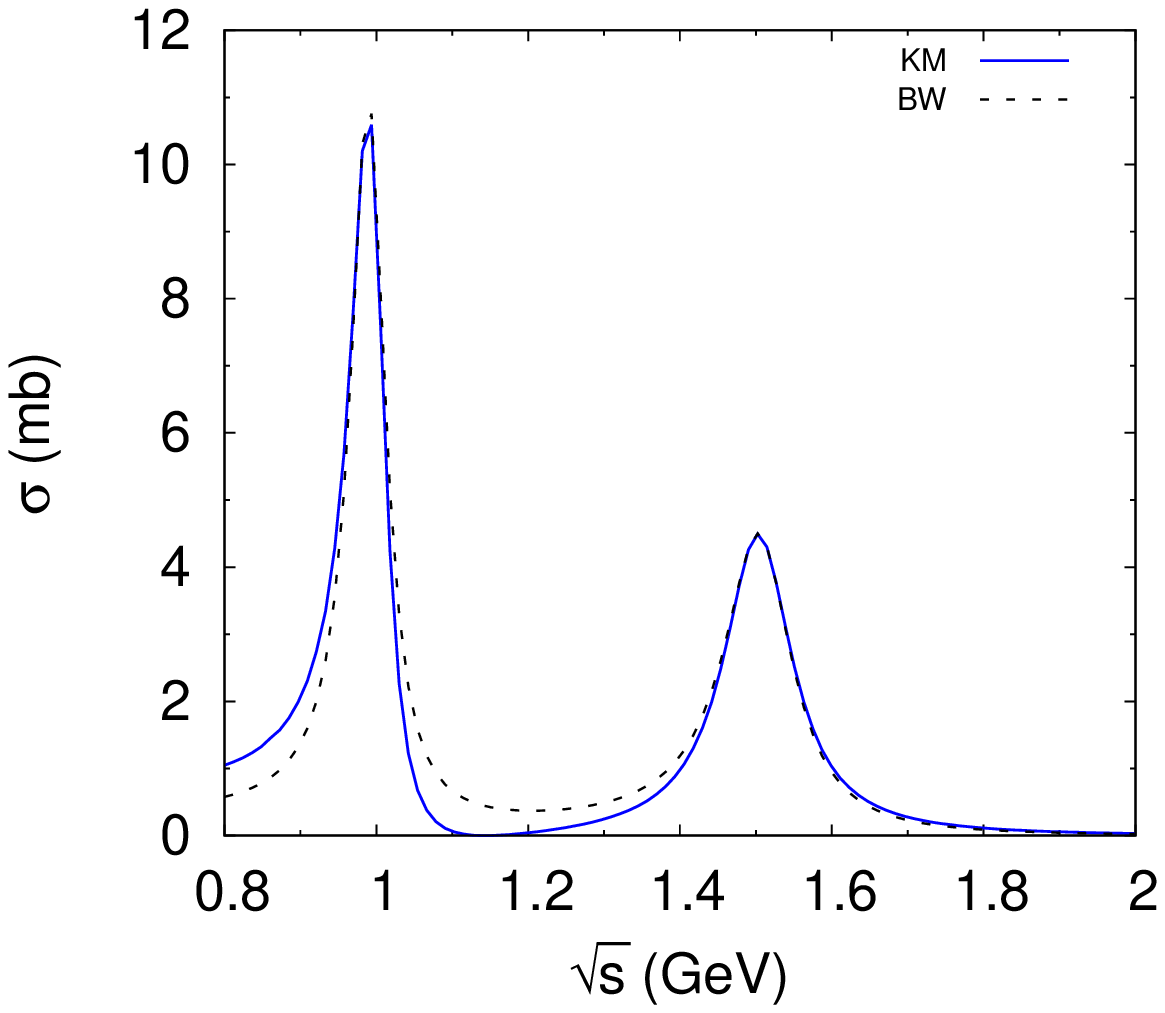}
\includegraphics[width=0.41\textwidth]{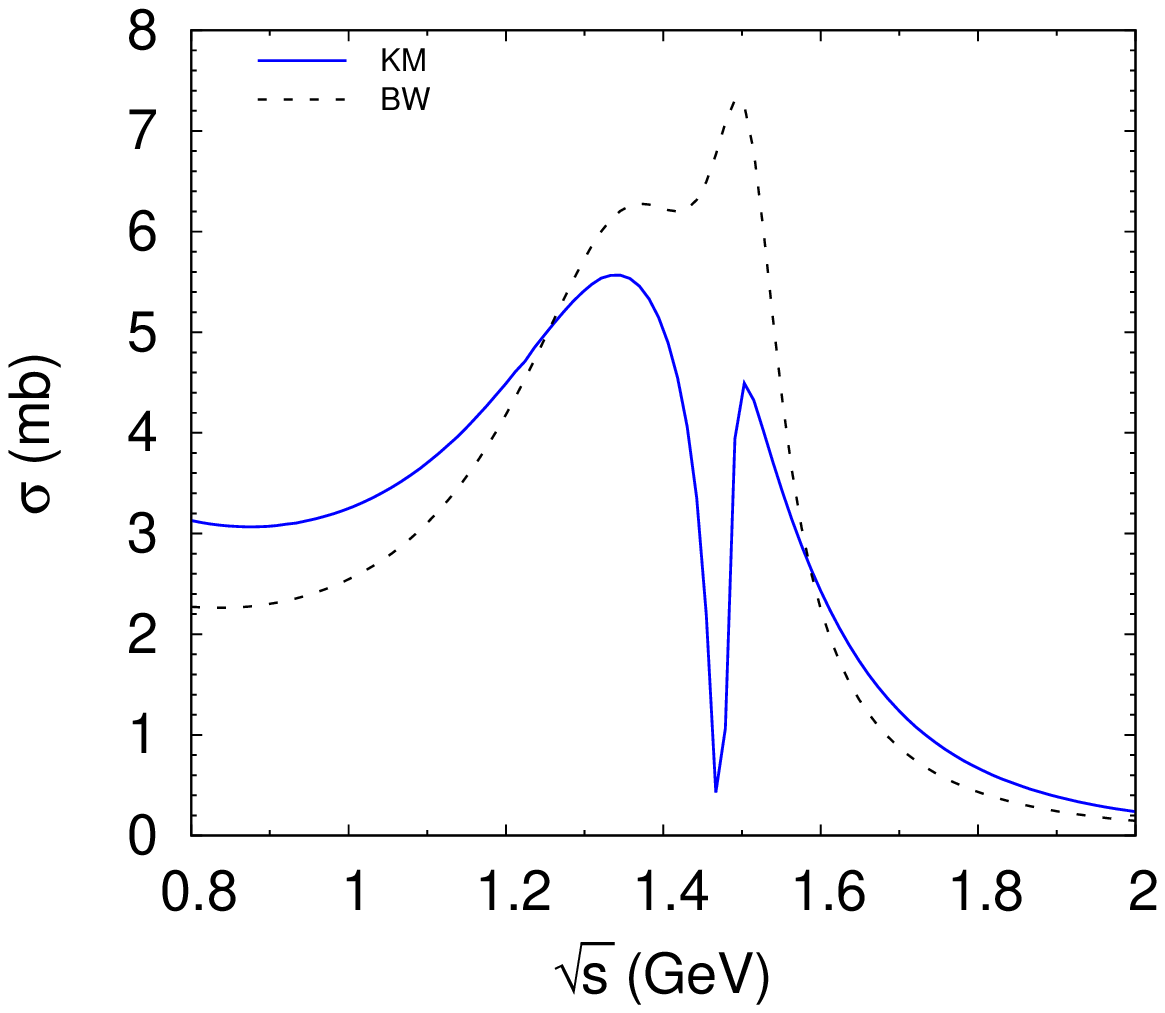}
\end{center}
\vspace{0.5cm}
 \caption{(Color online) 
 The variation of total cross section as a function of center of mass energy.
 Left panel shows total cross section of separated resonances $f_0(980)$ and $f_0(1500)$;
 right panel shows total cross section of overlapping resonances $f_0(1370)$ and $f_0(1500)$.
 The calculations using K-matrix formalism are shown
 using solid blue line (KM). Calculations using Breit-Wigner parametrization are shown using 
 dashed black line (BW).}
\label{Fig:ComparisionKMBW}
\end{figure*}

\section{The K-matrix formalism}\label{Sc: K-matrix}
The K-matrix formalism elegantly expresses the unitarity of the S-matrix for the processes of type $ab\rightarrow cd$, where $a$, $b$ and $c$, $d$ are hadrons. We provide only
a brief summary of the formalism in this section, interested readers are referred to Refs. \cite{Chung:1995dx,martin1970elementary,Wiranata:2013oaa}.\par

In general, the amplitude that an initial state $|i\rangle$ to be scattered to the final state $|f\rangle$ is
\begin{equation}
 S_{fi}=\langle f|S| i\rangle,
\end{equation}
where $S$ is called the scattering operator. Splitting the probability of non-interaction 
$I$ and interaction by defining the transition operator $T$, 
we have
\begin{equation}\label{Eq:TMatrix}
 S=I+2iT,
\end{equation}
where $I$ is the identity operator. Conservation of probability implies that scattering matrix $S$ should be unitary, i.e.,
\begin{equation}
 SS^{\dagger}=S^{\dagger}S=I.
\end{equation}
From the unitarity of $S$, one gets
\begin{equation}
 T-T^{\dagger}=2iT^{\dagger}T=2iTT^{\dagger}.
\end{equation}
One may further rearrange this expression into
\begin{equation}
{(T^{-1}+iI)}^{\dagger}=T^{-1}+iI.
\end{equation}
Let us introduce the Hermitian operator $K$ via
\begin{equation}\label{Eq:KTReln}
 K^{-1}=T^{-1}+iI.
\end{equation}
Since the operator $K$ is Hermitian, the matrix is symmetric and the the eigenvalues are real. 
One can rewrite the components of $T$ matrix in terms of $K$ matrix as
\begin{eqnarray}\label{Eq:RealIMParts}
 \operatorname{Re} T&=&{(I+K^2)}^{-1}K=K{(I+K^2)}^{-1}\nonumber,\\
 \operatorname{Im} T&=&{(I+K^2)}^{-1}K^2=K^2{(I+K^2)}^{-1}.
\end{eqnarray}
Resonances appear as sum of poles in the K-matrix as
\begin{equation}\label{Eq:KmatrixkeyEqn}
 K_{ab\rightarrow cd}=\sum_{R}\frac{g_{R\rightarrow ab}(\sqrt{s}) g_{R\rightarrow cd}(\sqrt{s})}{m_R^2-s},
\end{equation}
where the sum on $R$ runs over the number of resonances with mass $m_R$, and the residue functions are given by
\begin{equation}
 g_{R\rightarrow ab}(\sqrt{s})=m_R\Gamma_{R\rightarrow ab}(\sqrt{s}),
\end{equation}
where $\sqrt{s}$ is the center of mass energy.
The energy dependent partial decay widths \cite{Chung:1995dx} are given by 
\begin{equation}\label{Eq:Width}
 \Gamma_{R\rightarrow ab}(\sqrt{s})=\Gamma^0_{R\rightarrow ab}(\sqrt{s})\frac{m_R}{\sqrt{s}}\frac{ q_{ab}}{q_{ab0}}{\left(B^l(q_{ab},q_{ab0})\right)}^2.
\end{equation}
The momentum $q_{ab}$ is given as
\begin{equation}\label{Eq:commomentum}
 q_{ab}(\sqrt{s})=\frac{1}{2\sqrt{s}}\sqrt{\left(s-{(m_a+m_b)}^2\right)\left(s-{(m_a-m_b)}^2\right)},
\end{equation}
 where $m_a$ and $m_b$ being the mass of decaying hadrons $a$ and $b$.

In Eq.($\ref{Eq:Width}$), $q_{ab0}=q_{ab}(m_R)$ is the resonance momentum at $\sqrt{s}=m_R$ and $\Gamma^0_R$ is the width of the pole at half maximum. The $B^l(q_{ab},q_{ab0})$
are the Blatt-Weisskopf barrier factors which can be expressed in terms of momentum $q_{ab}$ and resonance momentum $q_{ab0}$
for the orbital angular momentum $l$ as
\begin{equation}
 B^l_{R\rightarrow ab}(q_{ab},q_{ab0})=\frac{F_l(q_{ab})}{F_l(q_{ab0})}.
\end{equation}
The barrier factors $F_l(q)$ can be obtained using the following definition:
\begin{equation}
 F_l(z)=\frac{|h_l^{(1)}(1)|}{|z h_l^{(1)}(z)|},
\end{equation}
where $h_l^{(1)}(z)$ are spherical Hankel functions of the first kind and $z={\left(q/q_R\right)}^2$, with $q_R=0.1973$ GeV corresponding
to $1$ fm.

\par
The scattering amplitude $f(\theta)$ can be expressed as
\begin{equation}
 f(\sqrt{s},\theta)=\frac{1}{q_{ab}}\sum_l (2l+1)T^lP_l(\cos \theta),
\end{equation}
in terms of the interaction matrix $T^l(s)$. Here $P_l(\cos \theta)$ are the Legendre polynomials for the angular 
momentum $l$ and $\theta$ is the center of mass scattering angle. The cross section for the process $ab\rightarrow cd$
can be given in terms of terms of scattering amplitude
\begin{equation}
 \sigma(\sqrt{s},\theta)={|f(\sqrt{s},\theta)|}^2.
\end{equation}

If we use partial decomposition of the $T$ matrix,
\begin{equation}
 T^l=e^{i\delta_l}\sin \delta_l,
\end{equation}
one can relate the phase shift in a single resonance of mass $m_1$ to the K-matrix using the relations in Eq.(\ref{Eq:RealIMParts}),
\begin{equation}
 K=\frac{m_1\Gamma_1(\sqrt{s})}{m_1^2-s}=\tan \delta_l.
\end{equation}

\subsection{Three body decay}\label{Sc: Three Body}
Let the resonance $R$ with mass $m_R$ decay into three other particles $a$, $b$ and $c$ of masses $m_a$, $m_b$ and $m_c$. The residue function is given by
\begin{equation}
 g_{R\rightarrow abc}(\sqrt{s})=\frac{1}{2m_{\pi}^2}\int d\phi_3{|\Gamma(\sqrt{s})|}^2,
\end{equation}
where $\phi_3$ is the three body Lorentz invariant phase space and we have scaled it by pion mass ($m_{\pi}$) to make it dimensionless. 
The three body phase space can be expressed as
\begin{align}
\begin{split}
 \phi_3 &=\int\frac{d^3p_1}{{(2{\pi})}^3}\frac{1}{2E_1}\frac{d^3p_2}{{(2{\pi})}^3}\frac{1}{2E_2}\times\\
& \frac{d^3p_3}{{(2{\pi})}^3}\frac{1}{2E_3}{(2{\pi})}^4{\delta}^4\left(p-\sum_{i=1}^3 p_i\right).\\
&=\frac{R_3(\sqrt{s})}{{(2\pi)}^5},
 \end{split}
\end{align}
where $E_i$'s and $p_i$'s are energies and the momenta of the decaying particles in the resonance rest frame. 
The function $R_3(\sqrt{s})$ is expressed as
\begin{equation}
 R_3(\sqrt{s})=\frac{{\pi}^2}{4s}\int_{s_2^{min}}^{s_2^{max}}\frac{ds_2}{s_2}\lambda^{\frac{1}{2}}
 (s_2,s,m_a^2)\lambda^{\frac{1}{2}}(s_2,m_b^2,m_c^2),
\end{equation}
where $s_2^{min}={\left(m_b+m_c\right)}^2$ and $s_2^{max}={\left(\sqrt{s}-m_a\right)}^2$ and the $\lambda$'s are the
Kallen functions \cite{Book_Byckling,Lo:2017sde}. They can be defined as
\begin{equation}
 \lambda(x,y,z)={(x-y-z)}^2-4yz.
\end{equation}
If we assume that the width $\Gamma(\sqrt{s})$ is a slowly varying function of energy, it can be pulled out of the integration sign and then finally
we have
\begin{equation}
 g_{R\rightarrow abc}(\sqrt{s})=\frac{1}{{(2\pi)}^5}\frac{R_3(\sqrt{s}){|\Gamma(\sqrt{s})|}^2}{2m_{\pi}^2}.
\end{equation}

\section{The Breit-Wigner parametrization}\label{Sc: BW}
The interaction matrix or the T matrix that was defined in Eq.(\ref{Eq:TMatrix}) for the relativistic single particle resonance can be parametrized in the Breit-Wigner form as \cite{Patrignani:2016xqp}
\begin{equation}\label{Eq:BWTMatix}
 T=\frac{m_R\Gamma_{R\rightarrow ab}(\sqrt{s})}{(m_R^2-s)-im_R\Gamma_R^{tot}(\sqrt{s})},
\end{equation}
where $\Gamma_{R}^{tot}=\sum_{i,j}\Gamma_{R\rightarrow ij}$ is the total width and $\Gamma_{R\rightarrow ij}$ is the partial width
for a given channel $R\rightarrow ij$ of the resonance $R$ respectively.\par

The cross section for an elastic scattering reaction $a+b\rightarrow R\rightarrow a+b$ is then given as,
\begin{equation}
 \sigma(\sqrt{s},\theta)=\frac{g_{I,l}}{q_{ab}^2}\frac{m_R^2\Gamma^2_{R\rightarrow ab}}{{(m_R^2-s)}^2+m^2_R{\Gamma_R^{tot}}^2}P_l(\cos \theta),
\end{equation}
where $g_{I,l}$ are the symmetry factors containing the isospin and spin multiplicities of the corresponding resonance $R$. The center of 
mass momentum $q_{ab}$ is the same as given in Eq.(\ref{Eq:commomentum}), $P_l(\cos \theta)$ are the Legendre polynomials for the angular 
momentum $l$ and $\theta$ is the center of mass scattering angle. The partial decay widths $\Gamma_{R\rightarrow ab}(\sqrt{s})$ are same as given in Eq.(\ref{Eq:Width}),

\section{Comparisons between K-matrix and Breit-Wigner approach}\label{Sc: Comparison}

Consider a $\pi\pi$ scattering at center of mass energy $\sqrt{s}$, which has two resonance with mass $m_1$ and $m_2$ coupling to a certain
channel $l$. From Eq.(\ref{Eq:KmatrixkeyEqn}) we have
\begin{equation}\label{Eq:KMTwo}
 K=\frac{m_1\Gamma_1(\sqrt{s})}{m_1^2-s}+\frac{m_2\Gamma_2(\sqrt{s})}{m_2^2-s},
\end{equation}
i.e the resonances are summed in the K-matrix. We can use Eq.(\ref{Eq:KTReln}) to get the T-matrix as
\begin{align}
 \begin{split}
  T = \frac{m_1\Gamma_{1}(\sqrt{s})}{(m_1^2-s)-im_1\Gamma_1(\sqrt{s})-i\frac{m_1^2-s}{m_2^2-s}m_2\Gamma_2(\sqrt{s})}+\\
  \frac{m_2\Gamma_{2}(\sqrt{s})}{(m_2^2-s)-im_2\Gamma_2(\sqrt{s})-i\frac{m_2^2-s}{m_1^2-s}m_1\Gamma_1(\sqrt{s})}.
  \label{Eq:TmatrixfromKmatrix}
 \end{split}
\end{align}
\par
If $m_1$ and $m_2$ are far apart relative to their widths, then $K$ is dominated either by $m_1$ or $m_2$ depending on whether $\sqrt{s}$ is
near $m_1$ or $m_2$. The transition amplitude is then given, using Eq.(\ref{Eq:TmatrixfromKmatrix}) approximately as the sum
\begin{equation}
 T \approx \frac{m_1\Gamma_{1}(\sqrt{s})}{(m_1^2-s)-im_1\Gamma_1(\sqrt{s})}+\frac{m_2\Gamma_{2}(\sqrt{s})}{(m_2^2-s)-im_2\Gamma_2(\sqrt{s})},
\end{equation}
which shows that the result is same as adding two Breit-Wigner forms Eq.(\ref{Eq:BWTMatix}) with mass $m_1$, $m_2$ and widths $\Gamma_1$, $\Gamma_2$. 
The left panel of Fig. \ref{Fig:ComparisionKMBW} compares the results of total cross-section in K-matrix and Breit-Wigner formalism for two separated
resonances $f_0(980)$ and $f_0(1500)$ of mass $m_1=990$ MeV, $\Gamma_1=55$ MeV and $m_2=1505$ MeV, $\Gamma_2=109$ MeV. The results are almost identical except that the peak in Breit-Wigner
formalism is slightly larger than K-matrix formalism.
\par

In the limit in which the two states have same masses, i.e. $m_c=m_1=m_2$, then the transition amplitude becomes
\begin{equation}
 T = \frac{m_c\left(\Gamma_{1}(\sqrt{s})+\Gamma_{2}(\sqrt{s})\right)}{(m_c^2-s)-im_c\left(\Gamma_{1}(\sqrt{s})+\Gamma_{2}(\sqrt{s})\right)},
\end{equation}
which shows that the result is a single Breit-Wigner form but its total width is now sum of the two individual widths. 
The right panel of Fig. \ref{Fig:ComparisionKMBW} compares the results of total cross-section in K-matrix and Breit-Wigner formalism for two overlapping
resonances $f_0(1370)$ and $f_0(1500)$ of mass $m_1=1370$ MeV, $\Gamma_1=350$ MeV and $m_2=1505$ MeV, $\Gamma_2=109$ MeV. The results shows that the Breit-Wigner 
parametrization overestimates the cross-section both at the peak and in the middle of the overlapping resonances. In such cases of two nearby 
resonances the Breit-Wigner form Eq.(\ref{Eq:BWTMatix}) is not strictly valid and the correct equation Eq.(\ref{Eq:KMTwo}) must be used.

\section{Relativistic Virial Expansion}\label{Sc: VE}
\begin{figure*}
\begin{center}
\includegraphics[width=0.32\textwidth]{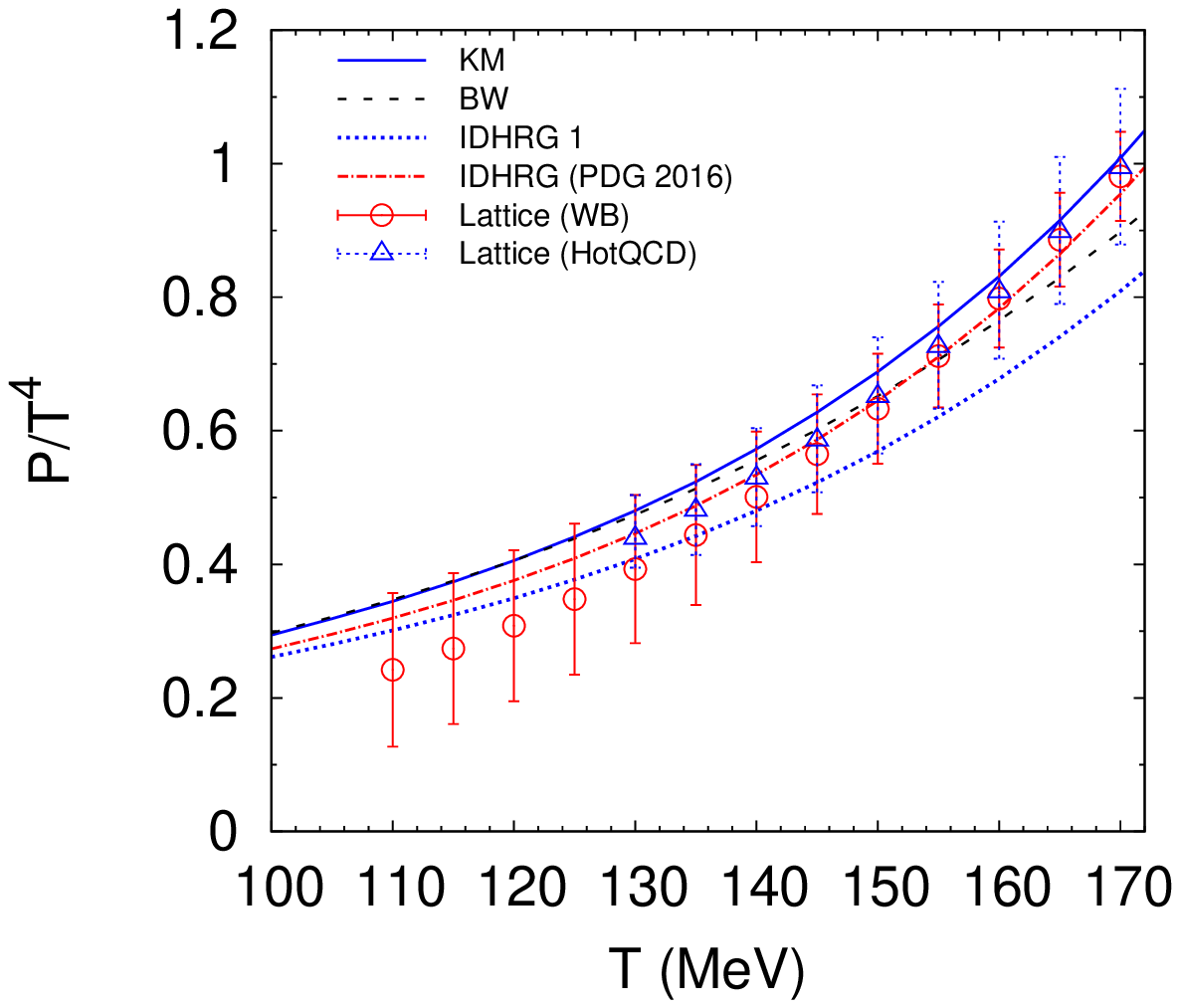}
\includegraphics[width=0.32\textwidth]{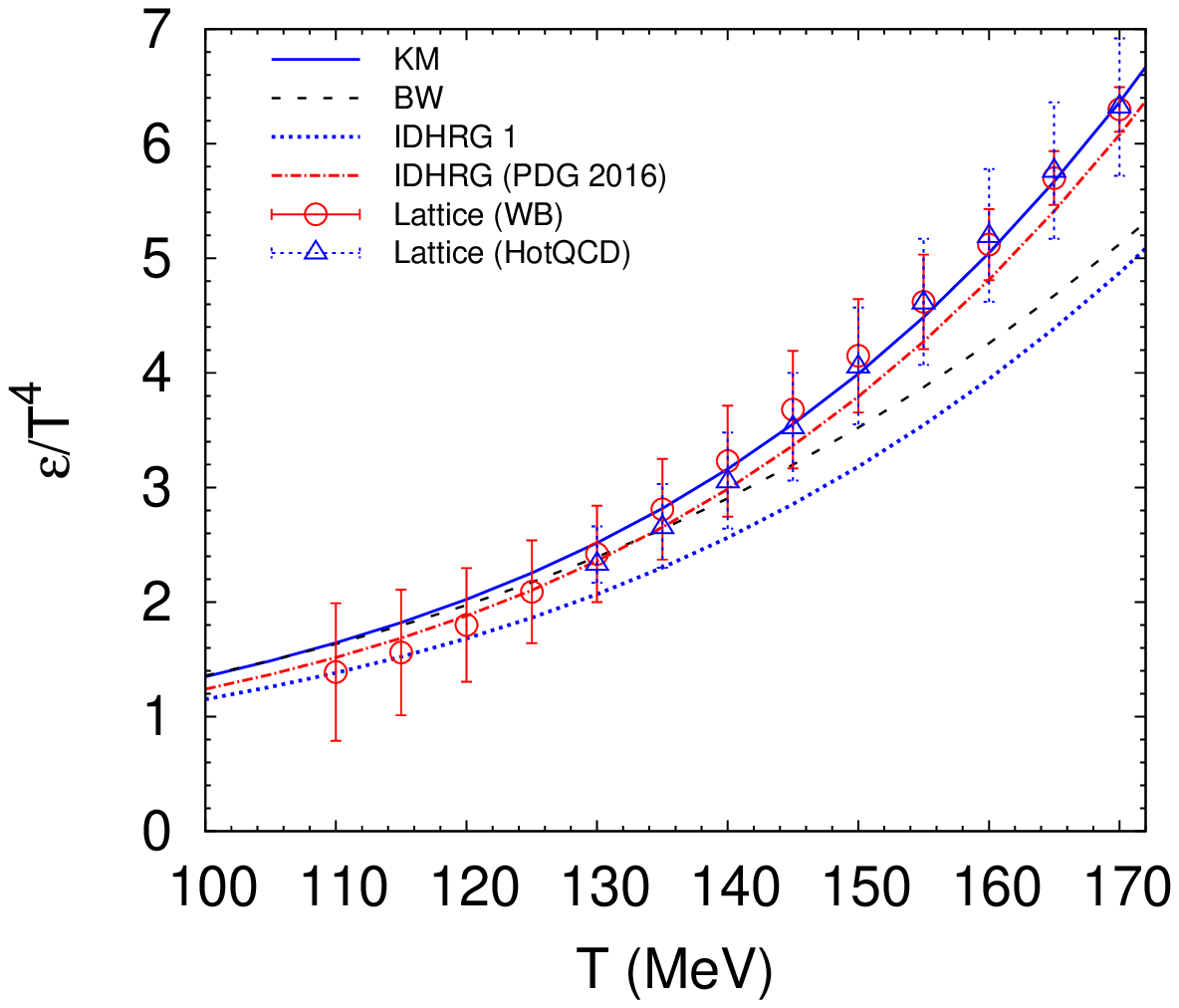}
\includegraphics[width=0.32\textwidth]{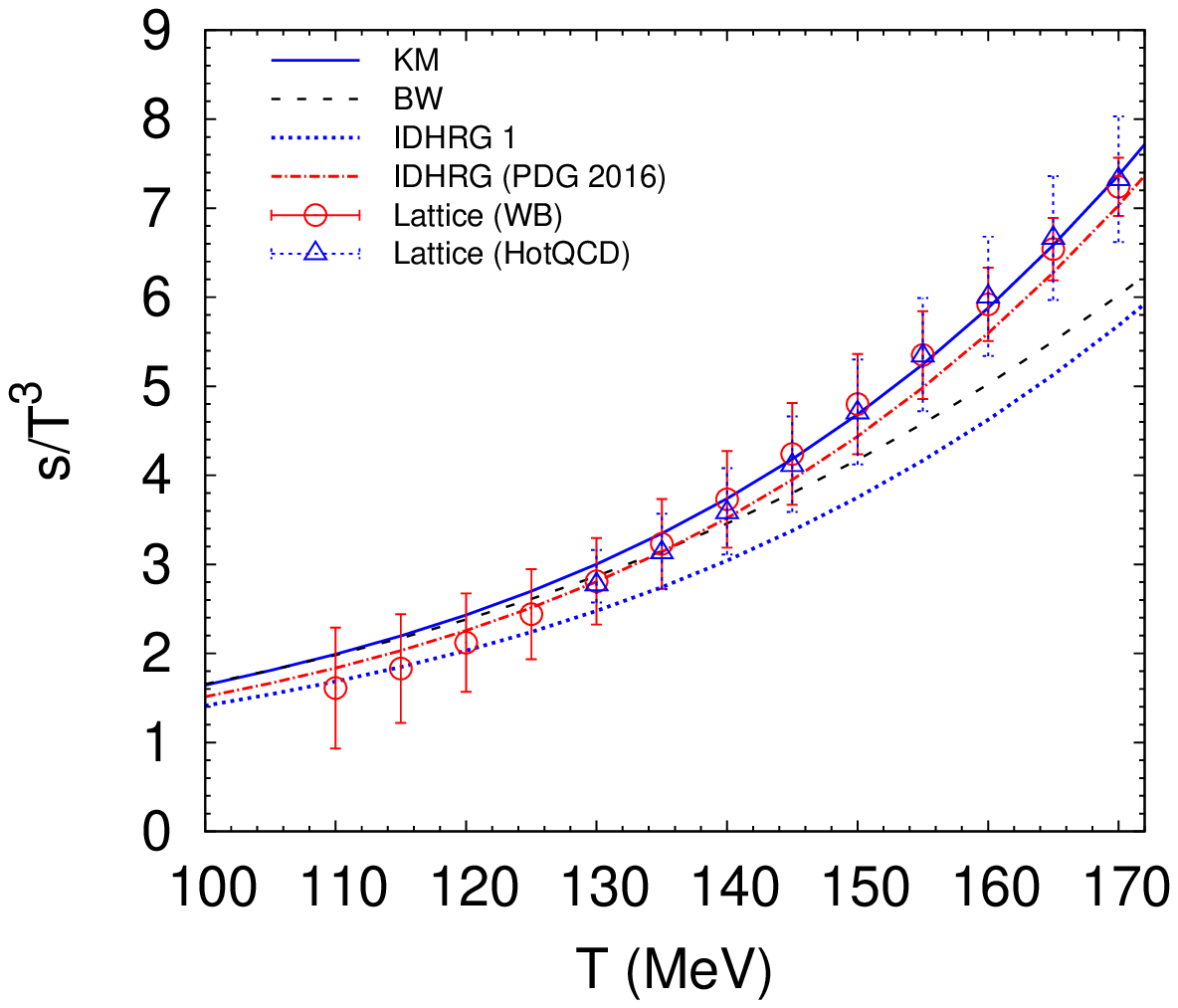}
\includegraphics[width=0.32\textwidth]{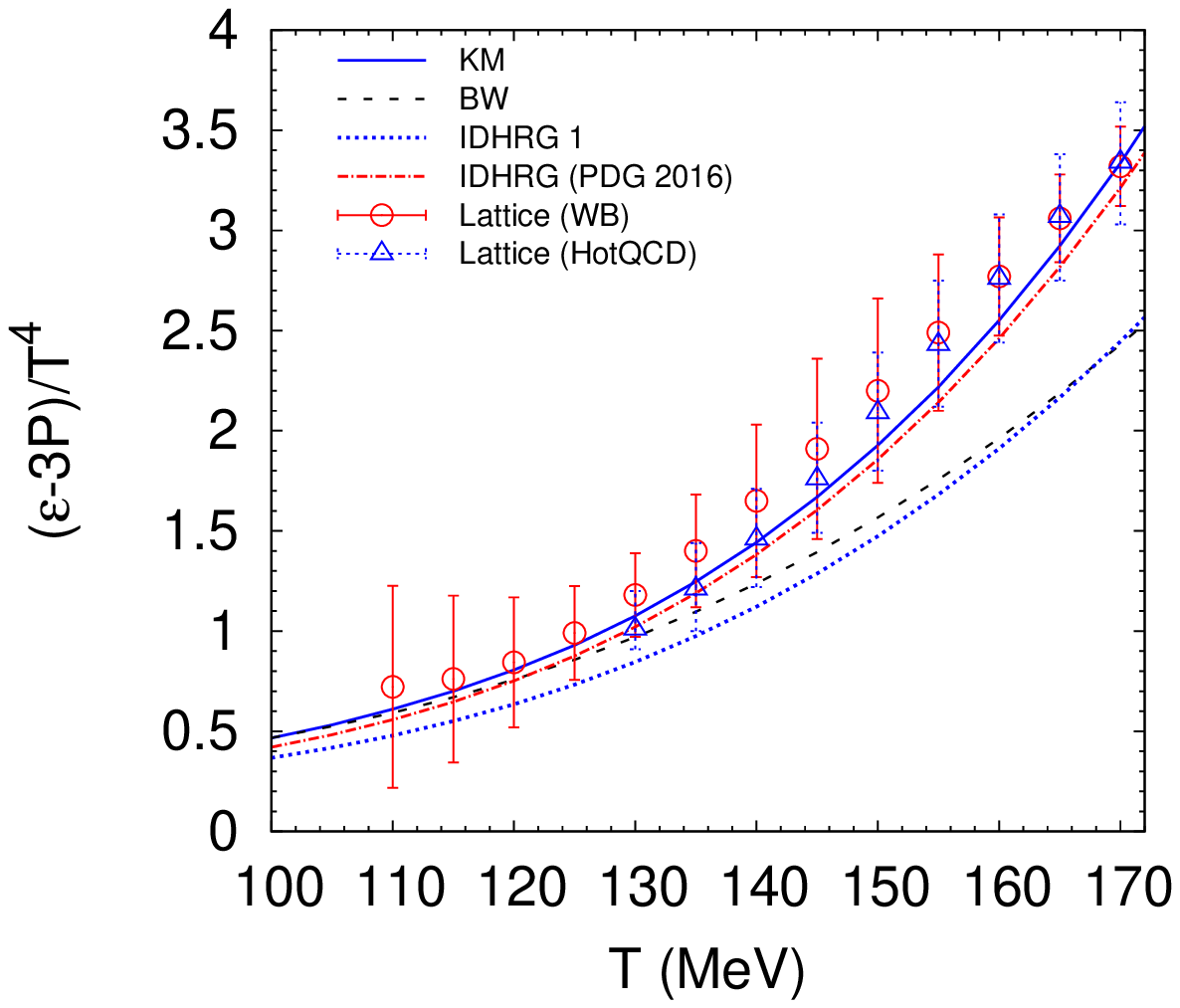}
\end{center}
\vspace{0.5cm}
 \caption{(Color online) Temperature dependence of various thermodynamic quantities at zero chemical potential. The calculations using K-matrix formalism are shown
 using solid blue line (KM). Calculations using Breit-Wigner parametrization are shown using dashed black line (BW). IDHRG 1 corresponds to results of 
 ideal HRG, with same number of particles as used in KM/BW parametrization whereas IDHRG (PDG 2016) includes all the hadrons and resonances 
 listed in PDG 2016 \cite{Patrignani:2016xqp}. Results are compared with lattice QCD data of Refs.~\cite{Borsanyi:2013bia} (WB) and 
\cite{Bazavov:2014pvz} (HotQCD).}
\label{fig:EOS}
\end{figure*}

The most natural way to incorporate interaction among a gas of hadrons is to use relativistic virial expansion introduced by Dashen \textit{et al} \cite{Dashen:1969ep}. 
The formalism allows one to compute the thermodynamics variables of a system in a grand canonical ensemble, once the S-matrix is known. In general, the logarithm of the partition
function can be written as
\begin{equation}\label{Eq:PartionFunction}
 \ln Z=\ln Z_0+\sum_{i_1,i_2}z_1^{i_1}z_2^{i_2}b(i_1,i_2),
\end{equation}
where $z_{1}$ and $z_{2}$ are fugacities of two species and $z=e^{\beta\mu}$. The chemical potential of $j$th particle is defined as 
$\mu_j=B_j\mu_B+S_j\mu_S+Q_j\mu_Q$ where $B_j$, $S_j$, $Q_j$ are baryon number, strangeness and electric charge and
$\mu$'s are the respective chemical potentials. The virial coefficients $b(i_1,i_2)$ are written as
\begin{align}\label{Eq:b2}
 b(i_1,i_2)=&\frac{V}{4\pi i}\int\frac{d^3p}{{(2\pi)}^3}\int d\varepsilon \exp\left(-\beta{(p^2+{\varepsilon}^2)}^{1/2}\right)\times\nonumber\\
 &{\left[A\left\{S^{-1}\frac{\partial S}{\partial \varepsilon}-\frac{\partial S^{-1}}{\partial \varepsilon}S\right\}\right]}_c.
\end{align}
In the above, the inverse temperature is denoted by $\beta$ while $V$, $p$ and $\varepsilon$ stand for the volume, the total center of mass momentum
and energy respectively. The labels $i_1$ and $i_2$ refer to channel of the S-matrix which has initial state containing $i_1+i_2$ particles. The 
symbol $A$ denotes the symmetrization (anti-symmetrization) operator for a system of bosons (fermions) while the subscript $c$ refers to trace 
over all linked diagrams. The lowest virial coefficient $b_2=b(i_1,i_2)/V$ as $V\rightarrow \infty$ corresponds to the case where $i_1=i_2=1$ and
in which the present study is mostly interested.\par

The S-matrix can be expressed in terms of phase shifts ${\delta}_l^I$ as \cite{Sakurai:2011zz}
\begin{equation}
 S(\varepsilon) = \sum_{l.I}(2l+1)(2I+1)\exp(2i{\delta}_l^I),
\end{equation}
where $l$ and $I$ denote angular momentum and isospin, respectively. On integrating Eq.(\ref{Eq:b2}) over the total momentum we have
\begin{equation}\label{Eq:Finalb2}
 b_2 = \frac{1}{2{\pi}^3\beta}\int_M^{\infty} d\varepsilon {\varepsilon}^2K_2(\beta \varepsilon)\sum_{l,I}{}^{'}g_{I,l}\frac{\partial {\delta}_l^I(\varepsilon) }{\partial \varepsilon}.
\end{equation}
The factor $g_{I,l} = (2I+1) (2l+1)$ is the degeneracy factor, $M$ is the invariant mass of the interacting pair at threshold and the factor 
$K_2(\beta \varepsilon)$ is the modified Bessel function of second kind. The prime over
the summation sign denotes that for given $l$ the sum over $I$ is restricted to values consistent with statistics.\par
Eq.(\ref{Eq:Finalb2}) shows 
that the contribution arising from interaction to thermodynamic variable, are in terms of phase shifts weighted by thermal factors. This factors give positive 
(attractive) or negative (repulsive) contribution depending on whether the derivative of phase shifts are positive or negative. The $b_2$ or alternatively phase shifts are
obtained from experiments or from theoretical calculations. In the present work, we determine the phase shifts from two different parametrization 
of the T-matrix: (i) K-matrix parametrization (ii) Breit-Wigner parametrization, which were discussed in section \ref{Sc: K-matrix} and 
\ref{Sc: BW}. Since, in this work we are interested only till the part corresponding to the second virial coefficient $b_2(\varepsilon)$ in the 
partition function Eq.(\ref{Eq:PartionFunction}), inserting Eq.(\ref{Eq:Finalb2}) into Eq.(\ref{Eq:PartionFunction}) one can immediately compute all the thermodynamic variables. We adopt the 
following relations from Ref. \cite{Venugopalan:1992hy}:

\begin{align}
\begin{split}
\operatorname{P_{int}}&=\frac{1}{\beta}\frac{\partial \ln Z_{int}}{\partial V}\\
&=\frac{z_1 z_2}{2{\pi}^3{\beta}^2}\int_{M}^{\infty} d\varepsilon {\varepsilon}^2K_2(\beta\varepsilon)\sum_{I,l}{}^{'}g_{I,l}
\frac{\partial {\delta}_l^I(\varepsilon)}{\partial \varepsilon},
 \label{interactingpressure}
\end{split}
 \end{align}
\begin{align}
\begin{split}
 \operatorname{\varepsilon_{int}}&=-\frac{1}{V}{\left(\frac{\partial \ln Z_{int}}{\partial \beta}\right)}_{z}\\
 &=\frac{z_1 z_2}{8{\pi}^3{\beta}}\int_{M}^{\infty} d\varepsilon {\varepsilon}^3\left[
 K_1(\beta\varepsilon)+3K_3(\beta\varepsilon)\right]\times\\
& \sum_{I,l}{}^{'}g_{I,l}\frac{\partial {\delta}_l^I(\varepsilon)}{\partial \varepsilon},
\label{interactingenegydensity}
\end{split}
\end{align}
\begin{align}
\begin{split}
\operatorname{s_{int}}&=-\frac{{\beta}^2}{V}{\left(\frac{\partial (T\ln Z_{int})}{\partial \beta}\right)}_{V,\mu}\\
&=\frac{z_1 z_2}{2{\pi}^3}\int_{M}^{\infty} d\varepsilon {\varepsilon}^3 K_3(\beta\varepsilon)\sum_{I,l}{}^{'}g_{I,l}\frac{\partial {\delta}_l^I(\varepsilon)}
{\partial \varepsilon}\\
&-{(\mu_1+\mu_2)}{\beta}^2\operatorname{P_{int}},
 \label{interactinentropydensity}
\end{split}
 \end{align}
 \begin{align}
 \operatorname{n_{int}}&= \frac{T}{V}{\left(\frac{\partial \ln Z_{int}}{\partial \mu}\right)}_{V,T}\\
 &= \frac{z_1 z_2}{{\pi}^3{\beta}}\int_{M}^{\infty} d\varepsilon {\varepsilon}^2K_2(\beta\varepsilon)\sum_{I,l}{}^{'}g_{I,l}
\frac{\partial {\delta}_l^I(\varepsilon)}{\partial \varepsilon},
 \label{interactingnumberdensity}
\end{align}
and the ideal gas counterpart can be obtained from the first term of Eq.(\ref{Eq:PartionFunction}) as follows: 
\begin{align}
\operatorname{P_{id}}=\sum_{h}\frac{g_h}{2{\pi}^2}m_h^2T^2\sum_{j=1}^{\infty}{(\pm 1)}^{j-1}(z^j/j^2)K_2(j\beta m_h),
\label{idealpressure}
\end{align}
\begin{align}
\operatorname{\varepsilon_{id}}=\sum_{h}\frac{g_h}{16{\pi}^2}m_h^4\sum_{j=1}^{\infty}{(\pm 1)}^{j-1}z^j
\left[K_4(j\beta m_h)-K_0(j\beta m_h)\right],
\label{idealenegydensity}
\end{align}
\begin{align}
\operatorname{n_{id}}=\sum_{h}\frac{g_h}{2{\pi}^2}m_h^2T\sum_{j=1}^{\infty}{(\pm 1)}^{j-1}(z^j/j)K_2(j\beta m_h),
\label{idealnumberdensity}
\end{align}
\begin{align}
\begin{split}
\hspace*{-3.5cm} \operatorname{s_{id}}=\beta(\operatorname{\varepsilon_{id}}+\operatorname{P_{id}}-\mu\operatorname{n_{id}}),
\end{split}
\end{align}
where $h$ denotes the stable hadron index. The total pressure of the system is the sum of ideal and interacting parts, i.e,
\begin{equation}
 P=\operatorname{P_{id}}+\operatorname{P_{int}}
\end{equation}
and subsequent relationships hold for other quantities.\par 

The susceptibilities of conserved charges can be calculated as \cite{Bellwied:2015lba}
\begin{equation}
 \chi_{BSQ}^{xyz}=\frac{\partial^{x+y+z}(P/T^4)}{\partial{(\mu_B/T)}^x\partial{(\mu_S/T)}^y\partial{(\mu_Q/T)}^z},
\end{equation}
where $x$, $y$ and $z$ are the order of derivatives of the quantities $B$, $S$ and $Q$. 
\begin{figure*}
\begin{center}
\includegraphics[width=0.32\textwidth]{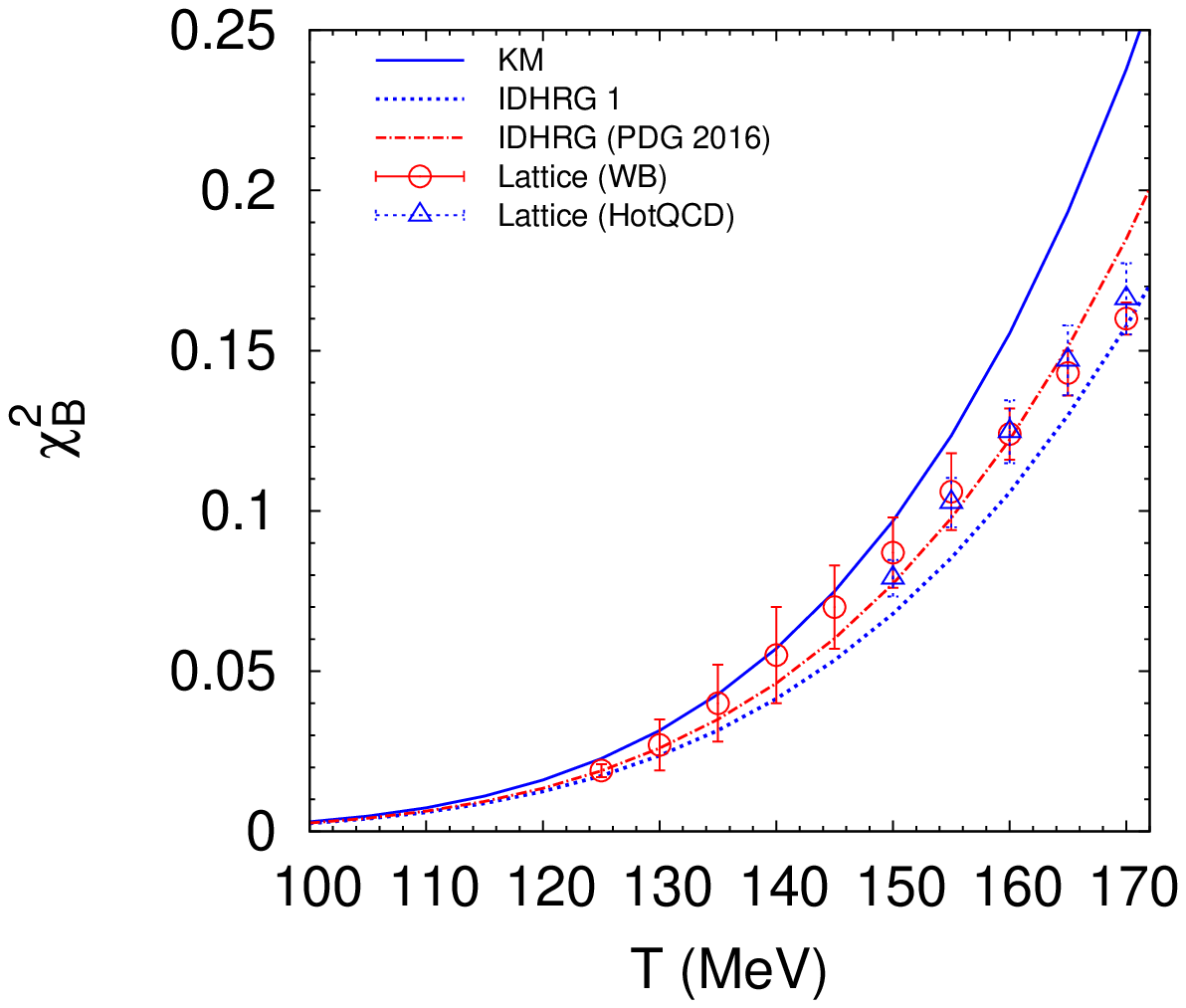}
\includegraphics[width=0.32\textwidth]{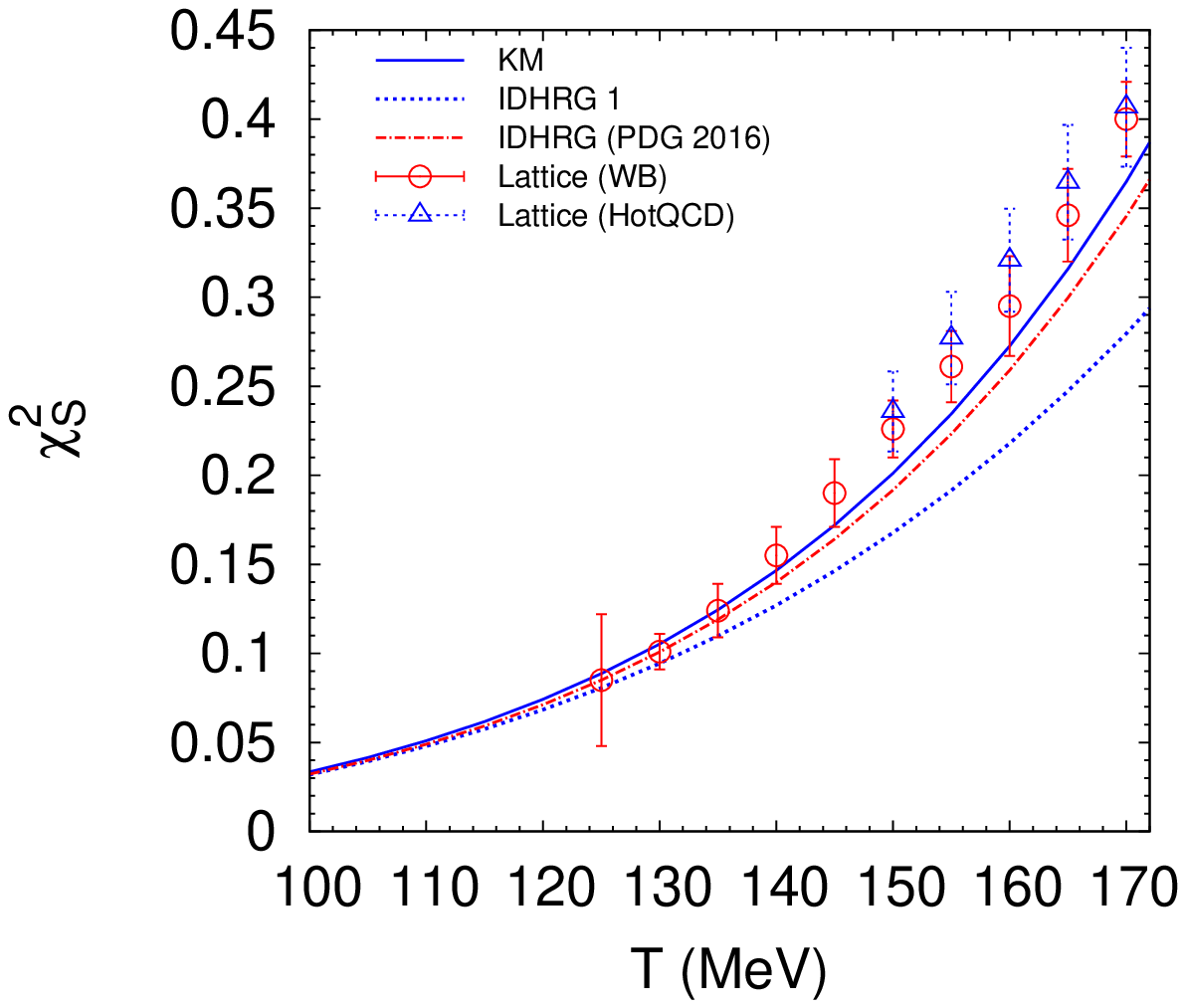}
\includegraphics[width=0.32\textwidth]{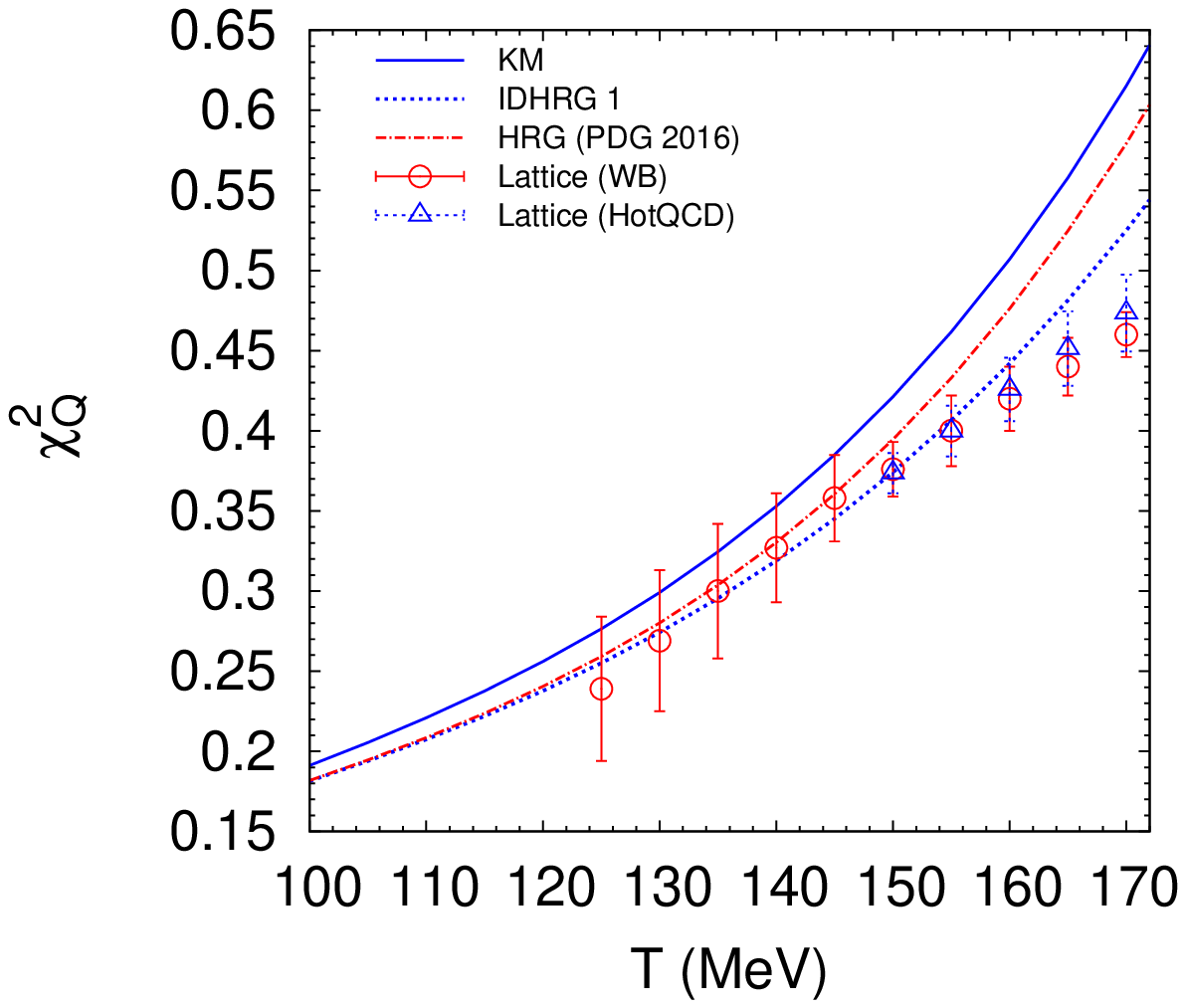}
\end{center}
\vspace{0.5cm}
 \caption{(Color online) Temperature dependence of second order diagonal susceptibilities at zero chemical potential. The calculations using K-matrix formalism are shown
 using solid blue line (KM). IDHRG 1 corresponds to results of 
 ideal HRG, with same number of particles as used in KM/BW parametrization whereas IDHRG (PDG 2016) includes all the hadrons and resonances 
 listed in PDG 2016 \cite{Patrignani:2016xqp}. Results are compared with lattice QCD data of Refs.~\cite{Borsanyi:2011sw} (WB)
 and~\cite{Bazavov:2012jq} (HotQCD).}
\label{fig:chidiagonal}
\end{figure*}

\begin{figure*}
\begin{center}
\includegraphics[width=0.32\textwidth]{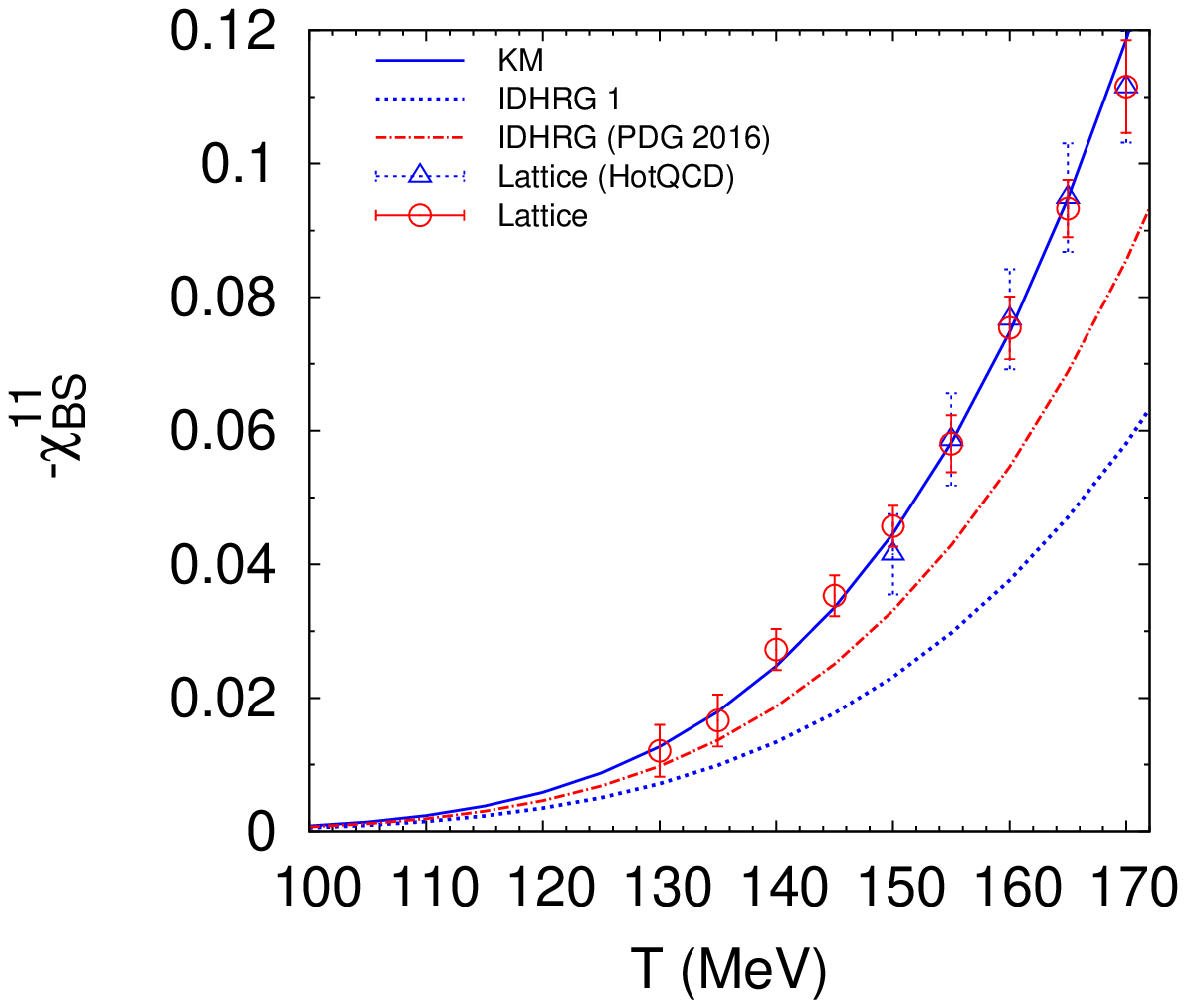}
\includegraphics[width=0.32\textwidth]{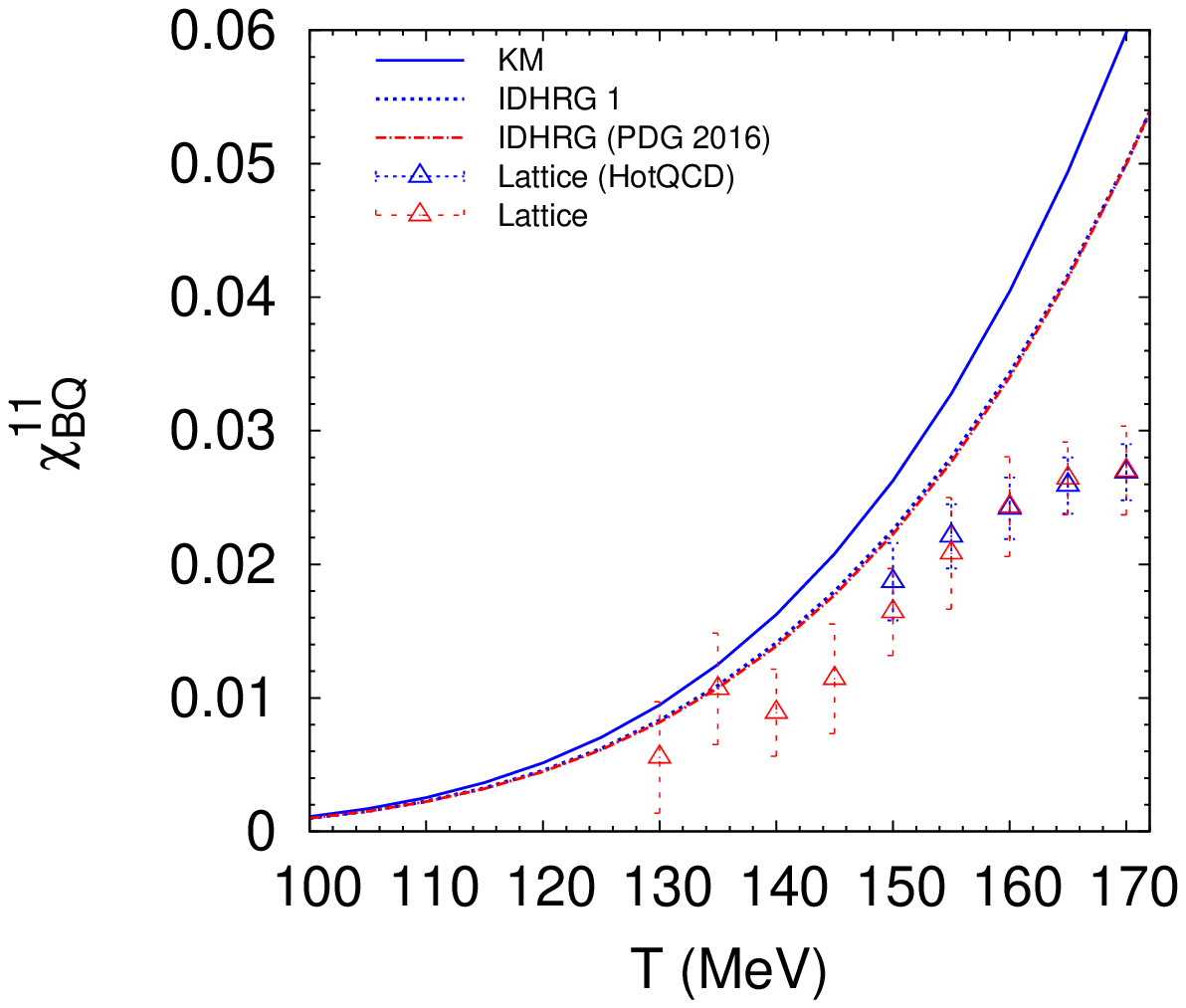}
\includegraphics[width=0.32\textwidth]{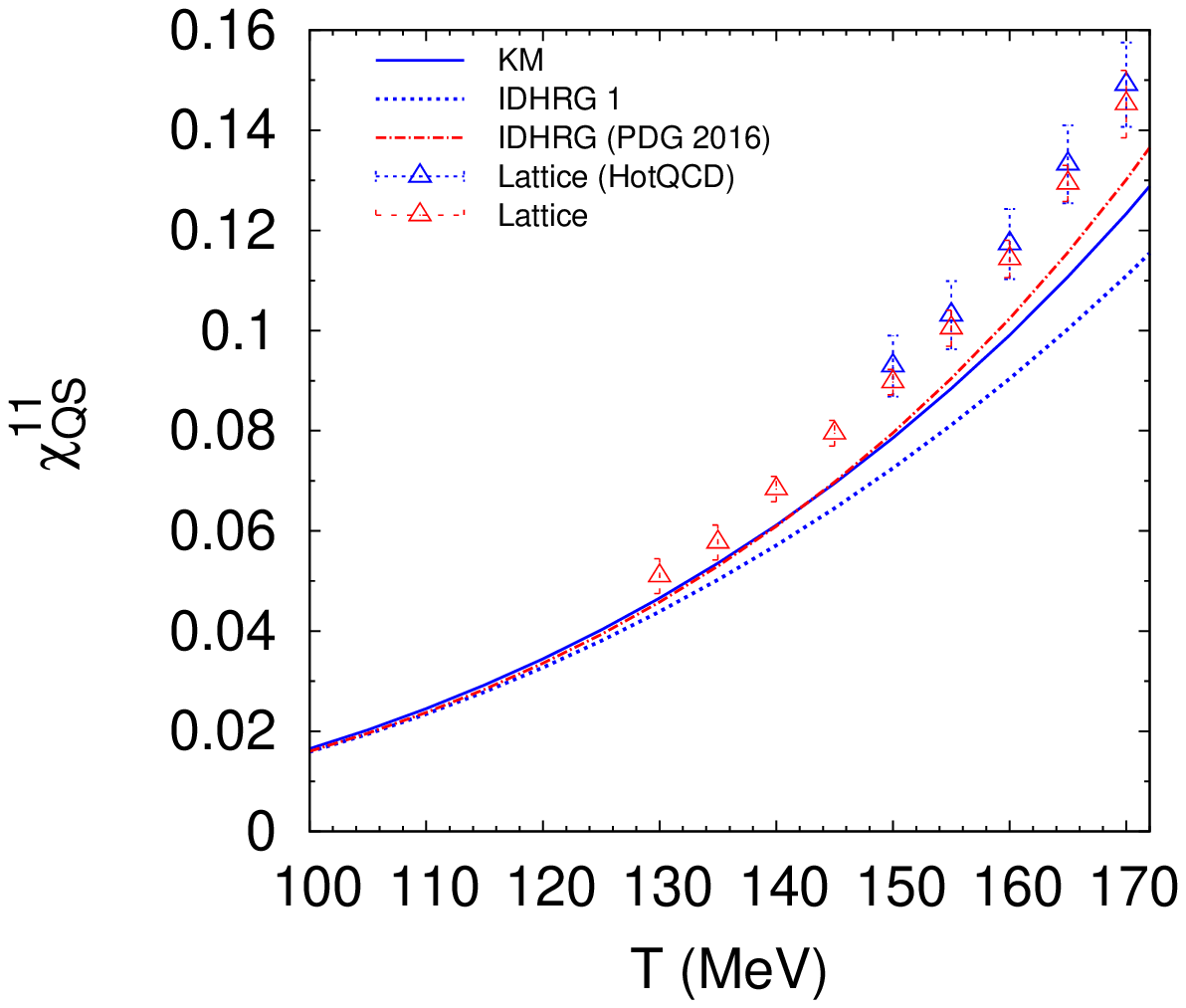}
\end{center}
\vspace{0.5cm}
 \caption{(Color online) Temperature dependence of second order off diagonal susceptibilities at zero chemical potential. The calculations using K-matrix formalism are shown
 using solid blue line (KM). IDHRG 1 corresponds to results of 
 ideal HRG, with same number of particles as used in KM/BW parametrization whereas IDHRG (PDG 2016) includes all the hadrons and resonances 
 listed in PDG 2016 \cite{Patrignani:2016xqp}. Results are compared with lattice QCD data of Refs.~\cite{Bazavov:2012jq} (HotQCD)
 and~\cite{Bellwied:2015lba} (Lattice).}
\label{fig:chioffdiagonal}
\end{figure*}\par

For a very narrow resonance of mass $m_R$, the phase shift $\delta^I_l$ changes rapidly through $\pi$ radians around $\varepsilon=m_R$ and can be approximated by a 
step function $\delta^I_l=\theta(\varepsilon-m_R)$. In such a limiting case $\partial\delta^I_l/\partial \varepsilon\approx \pi\delta(\varepsilon-m_R)$, then 
from  Eq.(\ref{Eq:Finalb2}) we have
\begin{equation}
 b_2=\frac{g_{I,l}}{2{\pi}^2}m_R^2TK_2(\beta m_R).
\end{equation}
This is similar to a stable hadron in the Maxwell-Boltzmann statistics.\par

A few points to be noted here. First, notice that the K-matrix and Breit-Wigner forms can only be able to take resonance interaction based on known mass and widths of the resonances and 
don't carry any information about repulsive channels which are known to exist from perturbative calculation \cite{Weinberg:1971mx} and experimental phase 
shifts \cite{Venugopalan:1992hy}. Recently some works \cite{Friman:2015zua} started looking at these repulsive channel but an adequate treatment of such channels is still missing. 
Second, apart form elastic interaction channel many inelastic channel can exist for a gas of interacting hadrons which the present study does not encompass
\cite{Huovinen:2016xxq}. Third, we have checked that the effects of three body interaction in equation of state is less than 5 \% in the range of temperature considered in this work.
Therefore we have not included effect of three body interaction in Sec. \ref{Sc: Results}.

\section{Results}\label{Sc: Results}
We have considered all the stable hadrons and resonances which have two body decay channels listed in PDG (2016) 
\cite{Patrignani:2016xqp}. In Fig. \ref{fig:EOS} we have shown temperature variation of different thermodynamic quantities such as, 
scaled pressure, energy density, entropy density and interaction measure $(\varepsilon-3P)/T^4$. Results of K-matrix 
parametrization are compared with Breit-Wigner parametrization and two variants of HRG models are shown. 
The IDHRG 1 considers the same number of resonances (i.e those decaying into two bodies) as the 
K-matrix/Breit-Wigner parametrization and the IDHRG (PDG 2016) considers all the resonances listed in PDG (2016). 
Later the results are compared with continuum extrapolated  LQCD data of Wuppertal-Budapest (WB) 
~\cite{Borsanyi:2013bia}
and Hot QCD collaboration ~\cite{Bazavov:2014pvz}. We notice that all thermodynamic quantities, 
calculated using K-matrix/Breit-Wigner
parametrization are larger compared to the results of IDHRG 1. This is because many resonances have finite width for example in the $\pi-\pi$ channel resonances like $f_0(500)$,
$\rho(770)$ or resonances
like $N(1440)$, $N(1520)$ in $\pi-N$ channel contribute substantially to the second virial coefficient. Differences in K-matrix and Breit-Wigner parametrization persist because 
of the presence of 
many overlapping resonances as mentioned in Sec. \ref{Sc: Comparison}. Particularly we found that the interaction measure, which is a measure of interactions in a medium, is 
well described in the K-matrix formalism. Thermodynamic quantities in IDHRG (PDG 2016) are larger compared to previous IDHRG 1 because of increased number of degeneracies. 
It is important to mention 
that the all our results are meaningful below the hadronic to quark gluon cross-over transition temperature
$T_c$ as predicted by LQCD \cite{Bazavov:2014pvz} which is around $145 - 163$ MeV. It must be noted that Ref. \cite{Wiranata:2013oaa} also used the K-matrix formalism to calculate interaction measure in an interacting
gas of $\pi-K-N-\eta$ and including the dominant resonances produced in two body elastic interaction.  However, their result underestimates the lattice results close to $T_c$, which has been improved in the present work 
by the inclusion of additional hadrons $\Xi-\Lambda-\Sigma$ and the corresponding resonances.
The above results show that our approach of using K-matrix formalism is in good agreement with LQCD.\par

The temperature dependencies of diagonal susceptibilities $\chi_B^2$, $\chi_S^2$, $\chi_Q^2$ are shown in Fig. \ref{fig:chidiagonal}. 
We compare our results with continuum extrapolated LQCD data of Refs.~\cite{Borsanyi:2011sw} (WB)
and~\cite{Bazavov:2012jq} (HotQCD).
Results of all diagonal susceptibilities in the K-matrix formalism is in better agreement with 
LQCD data, especially $\chi_S^2$,  up to cross-over temperature 
than IDHRG 1. However, it should be noted that $\chi_B^2$ and $\chi_Q^2$ in IDHRG (PDG 2016) also 
agrees with LQCD data but this is due to the increase in the number of 
degeneracies as mentioned earlier. Results of Breit-Wigner parameterization are not shown in the comparison because of its inherent inadequacy in treating in multiple resonances which leads to violation 
of unitarity\par

The temperature dependencies of off-diagonal susceptibilities $\chi^{11}_{BS}$, $\chi^{11}_{BQ}$, $\chi^{11}_{QS}$ are shown in Fig. \ref{fig:chioffdiagonal}. 
We compare our results with the continuum extrapolated LQCD data of Refs.~\cite{Bazavov:2012jq} (HotQCD)
 and~\cite{Bellwied:2015lba} (Lattice).
We have found that K-matrix formalism agrees with lattice data
for $\chi^{11}_{BS}$ 
but not for $\chi^{11}_{BQ}$ and $\chi^{11}_{QS}$. We think this might happen, since it is known \cite{Venugopalan:1992hy}
that many channels, mostly in the $N-N$ channels have dominant repulsive channels which could negate the influence of positive phase shifts, thereby contributing to the correlations.     

\section{Conclusion}\label{Sc: Conclusion}

To summarize we have included interaction properly in hadron resonance gas model using quantum virial expansion approach. 
The thermodynamic quantities were calculated by parameterizing the 
two body phase shifts using K-matrix formalism which preserves the
unitarity of S-matrix.  A good agreement with lattice QCD calculations is found for the equation of state using the above
formalism. Specifically we found that the interaction measure
($(\varepsilon-3P)/T^4$) as a function of temperature is well described in the K-matrix formalism and has been improved compared to the previous studies in Ref. \cite{Wiranata:2013oaa},
by the inclusion of additional hadrons. We found that IDHRG 1 (considering those resonances that decay into two stable hadrons) underestimates the lattice data for all the thermodynamic variables. 
However, IDHRG (PDG 2016) matches lattice data because of the increased number of degeneracies. Additionally we have calculated the diagonal and off-diagonal susceptibilities of conserved charges in the K-matrix formalism. 
The results of susceptibilities calculated in K-matrix formalism resembles lattice data quite well, especially in the strangeness sector below the cross-over region. However, observables
$\chi_{BQ}^{11}$ and $\chi_{QS}^{11}$ are not described satisfactorily in the present work.
This could be improved by incorporating inelastic collisions and repulsive interactions and it would be interesting to
introduce them in a future work.

\section*{Acknowledgement}
AD and SS would like to thank Victor Roy and Amaresh Jaiswal for helpful comments and discussions. BM acknowledges financial support from J C Bose National Fellowship of DST, Government
of India. AD and SS acknowledges financial support from DAE, Government of India.
%
\bibliographystyle{plain}

\begin{thebibliography}{99}

\bibitem{Aoki:2006we}
  Y.~Aoki, G.~Endrodi, Z.~Fodor, S.~D.~Katz and K.~K.~Szabo,
  Nature {\bf 443}, 675 (2006).

\bibitem{Borsanyi:2011sw} 
  S.~Borsanyi, Z.~Fodor, S.~D.~Katz, S.~Krieg, C.~Ratti and K.~Szabo,
  JHEP {\bf 1201}, 138 (2012).
  
\bibitem{Bazavov:2012jq} 
  A.~Bazavov {\it et al.} [HotQCD Collaboration],
  Phys.\ Rev.\ D {\bf 86}, 034509 (2012).
  
  
  \bibitem{Borsanyi:2013bia} 
  S.~Borsanyi, Z.~Fodor, C.~Hoelbling, S.~D.~Katz, S.~Krieg and K.~K.~Szabo,
  Phys.\ Lett.\ B {\bf 730}, 99 (2014).
  
  

  \bibitem{Bellwied:2013cta} 
  R.~Bellwied, S.~Borsanyi, Z.~Fodor, S.~D.~Katz and C.~Ratti,
  Phys.\ Rev.\ Lett.\  {\bf 111}, 202302 (2013).
 
\bibitem{Bazavov:2014pvz} 
  A.~Bazavov {\it et al.} [HotQCD Collaboration],
  Phys.\ Rev.\ D {\bf 90}, 094503 (2014).
  
  
  
\bibitem{Bellwied:2015lba} 
  R.~Bellwied, S.~Borsanyi, Z.~Fodor, S.~D.~Katz, A.~Pasztor, C.~Ratti and K.~K.~Szabo,
  Phys.\ Rev.\ D {\bf 92}, 114505 (2015).
  
 
  \bibitem{Gupta:2011wh} 
  S.~Gupta, X.~Luo, B.~Mohanty, H.~G.~Ritter and N.~Xu,
  Science {\bf 332}, 1525 (2011).  
  
  \bibitem{Asakawa:1989bq}
  M.~Asakawa and K.~Yazaki,
  Nucl.\ Phys.\ A {\bf 504}, 668 (1989).
  
  
    
\bibitem{Abelev:2009bw} 
  B.~I.~Abelev {\it et al.} [STAR Collaboration],
  Phys.\ Rev.\ C {\bf 81}, 024911 (2010).  
  
\bibitem{Adamczyk:2013dal} 
  L.~Adamczyk {\it et al.} [STAR Collaboration],
  Phys.\ Rev.\ Lett.\  {\bf 112}, 032302 (2014).  
  
 \bibitem{Agakishiev:2015bwu} 
  G.~Agakishiev {\it et al.} [HADES Collaboration],
  Eur.\ Phys.\ J.\ A {\bf 52}, 178 (2016).
  
  \bibitem{Ablyazimov:2017guv} 
  T.~Ablyazimov {\it et al.} [CBM Collaboration],
  Eur.\ Phys.\ J.\ A {\bf 53}, 60 (2017).
  
  \bibitem{Kekelidze:2017ual} 
  V.~D.~Kekelidze [NICA Collaboration],
  JINST {\bf 12}, C06012 (2017).
  
  

\bibitem{Hagedorn:1980kb}
  R.~Hagedorn and J.~Rafelski,
  Phys.\ Lett.\ B {\bf 97}, 136 (1980).
  
\bibitem{Rischke:1991ke}
  D.~H.~Rischke, M.~I.~Gorenstein, H.~Stoecker and W.~Greiner,
  Z.\ Phys.\ C {\bf 51}, 485 (1991).



\bibitem{Cleymans:1992jz}
  J.~Cleymans, M.~I.~Gorenstein, J.~Stalnacke and E.~Suhonen,
  Phys.\ Scripta {\bf 48}, 277 (1993).
  
  
\bibitem{BraunMunzinger:1994xr} 
  P.~Braun-Munzinger, J.~Stachel, J.~P.~Wessels and N.~Xu,
  Phys.\ Lett.\ B {\bf 344}, 43 (1995).
  
  
\bibitem{Cleymans:1996cd}
  J.~Cleymans, D.~Elliott, H.~Satz and R.~L.~Thews,
  Z.\ Phys.\ C {\bf 74}, 319 (1997).
  
\bibitem{Yen:1997rv}
  G.~D.~Yen, M.~I.~Gorenstein, W.~Greiner and S.~N.~Yang,
  Phys.\ Rev.\ C {\bf 56}, 2210 (1997).
  


\bibitem{BraunMunzinger:1999qy}
  P.~Braun-Munzinger, I.~Heppe and J.~Stachel,
  Phys.\ Lett.\ B {\bf 465}, 15 (1999).


\bibitem{Cleymans:1999st}
  J.~Cleymans and K.~Redlich,
  Phys.\ Rev.\ C {\bf 60}, 054908 (1999).



\bibitem{BraunMunzinger:2001ip}
  P.~Braun-Munzinger, D.~Magestro, K.~Redlich and J.~Stachel,
  Phys.\ Lett.\ B {\bf 518}, 41 (2001).

\bibitem{BraunMunzinger:2003zd}
  P.~Braun-Munzinger, K.~Redlich and J.~Stachel,
  invited review in {\it Quark Gluon Plasma 3}, edited by
R.C. Hwa and X.N. Wang, (World Scientific Publishing, 2004), [nucl-th/0304013].

 

\bibitem{Karsch:2003zq}
  F.~Karsch, K.~Redlich and A.~Tawfik,
  Phys.\ Lett.\ B {\bf 571}, 67 (2003).



\bibitem{Tawfik:2004sw}
  A.~Tawfik,
  Phys.\ Rev.\ D {\bf 71}, 054502 (2005).


\bibitem{Becattini:2005xt}
  F.~Becattini, J.~Manninen and M.~Gazdzicki,
  Phys.\ Rev.\ C {\bf 73}, 044905 (2006).


\bibitem{Andronic:2005yp}
  A.~Andronic, P.~Braun-Munzinger and J.~Stachel,
  Nucl.\ Phys.\ A {\bf 772}, 167 (2006).
  
  
\bibitem{Andronic:2008gu}
  A.~Andronic, P.~Braun-Munzinger and J.~Stachel,
  Phys.\ Lett.\ B {\bf 673}, 142 (2009).

\bibitem{Begun:2012rf}
  V.~V.~Begun, M.~Gazdzicki and M.~I.~Gorenstein,
  Phys.\ Rev.\ C {\bf 88}, 024902 (2013).
  
  
\bibitem{Andronic:2012ut}
  A.~Andronic, P.~Braun-Munzinger, J.~Stachel and M.~Winn,
  Phys.\ Lett.\ B {\bf 718}, 80 (2012).

\bibitem{Tiwari:2011km}
  S.~K.~Tiwari, P.~K.~Srivastava and C.~P.~Singh,
  Phys.\ Rev.\ C {\bf 85}, 014908 (2012).

\bibitem{Fu:2013gga}
  J.~Fu,
  Phys.\ Lett.\ B {\bf 722}, 144 (2013).
  
\bibitem{Tawfik:2013eua}
  A.~Tawfik,
  Phys.\ Rev.\ C {\bf 88}, 035203 (2013).
  
\bibitem{Bhattacharyya:2013oya}
  A.~Bhattacharyya, S.~Das, S.~K.~Ghosh, R.~Ray and S.~Samanta,
  Phys.\ Rev.\ C {\bf 90}, 034909 (2014).
  
\bibitem{Garg:2013ata}
  P.~Garg, D.~K.~Mishra, P.~K.~Netrakanti, B.~Mohanty, A.~K.~Mohanty, B.~K.~Singh and N.~Xu,
  Phys.\ Lett.\ B {\bf 726}, 691 (2013).


\bibitem{Bhattacharyya:2015zka}
  A.~Bhattacharyya, R.~Ray, S.~Samanta and S.~Sur,
  Phys.\ Rev.\ C {\bf 91}, 041901(R) (2015).
  
  \bibitem{Chatterjee:2013yga} 
  S.~Chatterjee, R.~M.~Godbole and S.~Gupta,
  Phys.\ Lett.\ B {\bf 727}, 554 (2013).
  
  \bibitem{Chatterjee:2014ysa} 
  S.~Chatterjee and B.~Mohanty,
  Phys.\ Rev.\ C {\bf 90}, 034908 (2014).
  
  \bibitem{Chatterjee:2014lfa} 
  S.~Chatterjee, B.~Mohanty and R.~Singh,
  Phys.\ Rev.\ C {\bf 92}, 024917 (2015).
  
  \bibitem{Becattini:2012xb} 
  F.~Becattini, M.~Bleicher, T.~Kollegger, T.~Schuster, J.~Steinheimer and R.~Stock,
  Phys.\ Rev.\ Lett.\  {\bf 111}, 082302 (2013).
  
  \bibitem{Bugaev:2013sfa} 
  K.~A.~Bugaev, D.~R.~Oliinychenko, J.~Cleymans, A.~I.~Ivanytskyi, I.~N.~Mishustin, E.~G.~Nikonov and V.~V.~Sagun,
  Europhys.\ Lett.\  {\bf 104}, 22002 (2013).
  
  \bibitem{Petran:2013lja} 
  M.~Petrán, J.~Letessier, V.~Petráček and J.~Rafelski,
  Phys.\ Rev.\ C {\bf 88}, 034907 (2013).
  
   \bibitem{Vovchenko:2014pka} 
  V.~Vovchenko, D.~V.~Anchishkin and M.~I.~Gorenstein,
  Phys.\ Rev.\ C {\bf 91}, 024905 (2015)
  
 
\bibitem{Kadam:2015xsa}
  G.~P.~Kadam and H.~Mishra,
  Phys.\ Rev.\ C {\bf 92}, 035203 (2015).

\bibitem{Kadam:2015fza}
  G.~P.~Kadam and H.~Mishra,
  Phys.\ Rev.\ C {\bf 93}, 025205 (2016).
  
\bibitem{Albright:2014gva}
  M.~Albright, J.~Kapusta and C.~Young,
  Phys.\ Rev.\ C {\bf 90}, 024915 (2014).
 
  

\bibitem{Albright:2015uua}
  M.~Albright, J.~Kapusta and C.~Young,
  Phys.\ Rev.\ C {\bf 92}, 044904 (2015).
  
   \bibitem{Bhattacharyya:2015pra}
  A.~Bhattacharyya, S.~K.~Ghosh, R.~Ray and S.~Samanta,
  Europhys.\ Lett.\  {\bf 115}, 62003 (2016).
 


  \bibitem{Kapusta:2016kpq}
  J.~Kapusta, M.~Albright and C.~Young,
  Eur.\ Phys.\ J.\ A {\bf 52}, 250 (2016).
  
\bibitem{Begun:2016cva}
  V.~Begun,
  Phys.\ Rev.\ C {\bf 94}, 054904 (2016).



\bibitem{Adak:2016jtk} 
  R.~P.~Adak, S.~Das, S.~K.~Ghosh, R.~Ray and S.~Samanta,
  Phys.\ Rev.\ C {\bf 96}, 014902 (2017).



\bibitem{Xu:2016skm}
  H.~j.~Xu,
  Phys.\ Lett.\ B {\bf 765}, 188 (2017).



\bibitem{Fu:2016baf} 
  J.~H.~Fu,
  Phys.\ Rev.\ C {\bf 96}, 034905 (2017).
  
  
\bibitem{Vovchenko:2015xja} 
  V.~Vovchenko, D.~V.~Anchishkin and M.~I.~Gorenstein,
  J.\ Phys.\ A {\bf 48}, 305001 (2015).



\bibitem{Vovchenko:2015vxa} 
  V.~Vovchenko, D.~V.~Anchishkin and M.~I.~Gorenstein,
  Phys.\ Rev.\ C {\bf 91}, 064314 (2015).


\bibitem{Vovchenko:2015pya} 
  V.~Vovchenko, D.~V.~Anchishkin, M.~I.~Gorenstein and R.~V.~Poberezhnyuk,
  Phys.\ Rev.\ C {\bf 92}, 054901 (2015).

  \bibitem{Broniowski:2015oha} 
  W.~Broniowski, F.~Giacosa and V.~Begun,
  Phys.\ Rev.\ C {\bf 92}, 034905 (2015).
  
   
  \bibitem{Vovchenko:2015idt} 
  V.~Vovchenko, V.~V.~Begun and M.~I.~Gorenstein,
  Phys.\ Rev.\ C {\bf 93}, 064906 (2016).
  
   
\bibitem{Redlich:2016dpb} 
  K.~Redlich and K.~Zalewski,
  Acta Phys.\ Polon.\ B {\bf 47}, 1943 (2016).


  
\bibitem{Vovchenko:2016rkn} 
  V.~Vovchenko, M.~I.~Gorenstein and H.~Stoecker,
  Phys.\ Rev.\ Lett.\  {\bf 118}, 182301 (2017).
  

\bibitem{Alba:2016fku} 
  P.~Alba, W.~M.~Alberico, A.~Nada, M.~Panero and H.~St\"{o}cker,
  Phys.\ Rev.\ D {\bf 95}, 094511 (2017).
  
\bibitem{Samanta:2017kmg} 
  S.~Samanta,
  arXiv:1702.01787 [hep-ph].
  
  \bibitem{Samanta:2017ohm} 
  S.~Samanta, S.~Ghosh and B.~Mohanty,
  arXiv:1706.07709 [hep-ph].
  
  \bibitem{Sarkar:2017ijd} 
  N.~Sarkar and P.~Ghosh,
  Phys.\ Rev.\ C {\bf 96}, 044901 (2017).
  
  \bibitem{Bhattacharyya:2017gwt} 
  A.~Bhattacharyya, S.~K.~Ghosh, S.~Maity, S.~Raha, R.~Ray, K.~Saha, S.~Samanta and S.~Upadhaya,
  arXiv:1708.04549 [hep-ph].
  
\bibitem{Chatterjee:2017yhp} 
  S.~Chatterjee, D.~Mishra, B.~Mohanty and S.~Samanta,
  Phys.\ Rev.\ C {\bf 96}, 054907 (2017).

   \bibitem{Alba:2016hwx} 
  P.~Alba, V.~Vovchenko, M.~I.~Gorenstein and H.~Stoecker,
  arXiv:1606.06542 [hep-ph].
  

 \bibitem{Samanta:2017yhh} 
  S.~Samanta and B.~Mohanty,
  Phys.\ Rev.\ C {\bf 97}, 015201 (2018). 
  
 
\bibitem{Alba:2017bbr}
  P.~Alba,
  arXiv:1711.02797 [nucl-th].
  
\bibitem{Olive:1980dy} 
  K.~A.~Olive,
  Nucl.\ Phys.\ B {\bf 190}, 483 (1981). 

  \bibitem{Olive:1982we} 
  K.~A.~Olive,
  Nucl.\ Phys.\ B {\bf 198}, 461 (1982).
 
  \bibitem{Dashen:1969ep} 
  R.~Dashen, S.~K.~Ma and H.~J.~Bernstein,
  Phys.\ Rev.\  {\bf 187}, 345 (1969).
  
  \bibitem{Weinhold:1997ig} 
  W.~Weinhold, B.~Friman and W.~Norenberg,
  Phys.\ Lett.\ B {\bf 433}, 236 (1998).
  \bibitem{Dobado:1998tv}
  A.~Dobado and J.~R.~Pelaez,
  Phys.\ Rev.\ D {\bf 59} (1999) 034004.
 
   
\bibitem{Venugopalan:1992hy} 
  R.~Venugopalan and M.~Prakash,
  Nucl.\ Phys.\ A {\bf 546}, 718 (1992).
  
\bibitem{Welke:1990za} 
  G.~M.~Welke, R.~Venugopalan and M.~Prakash,
  Phys.\ Lett.\ B {\bf 245}, 137 (1990).

  \bibitem{Friman:2015zua} 
  B.~Friman, P.~M.~Lo, M.~Marczenko, K.~Redlich and C.~Sasaki,
  Phys.\ Rev.\ D {\bf 92}, 074003 (2015).
  
   \bibitem{Huovinen:2016xxq} 
  P.~Huovinen, P.~M.~Lo, M.~Marczenko, K.~Morita, K.~Redlich and C.~Sasaki,
  Phys.\ Lett.\ B {\bf 769}, 509 (2017).
  

\bibitem{Wiranata:2013oaa} 
  A.~Wiranata, V.~Koch, M.~Prakash and X.~N.~Wang,
  Phys.\ Rev.\ C {\bf 88}, 044917 (2013).
  
\bibitem{Badalian:1981xj} 
  A.~M.~Badalian, L.~P.~Kok, M.~I.~Polikarpov and Y.~A.~Simonov,
  Phys.\ Rept.\  {\bf 82}, 31 (1982).
  
\bibitem{Brown:1991en} 
  G.~E.~Brown, J.~Stachel and G.~M.~Welke,
  Phys.\ Lett.\ B {\bf 253}, 19 (1991).
  
\bibitem{Sollfrank:1990qz} 
  J.~Sollfrank, P.~Koch and U.~W.~Heinz,
  Phys.\ Lett.\ B {\bf 252}, 256 (1990).

\bibitem{Gorenstein:1987zm} 
  M.~I.~Gorenstein, M.~S.~Tsai and S.~N.~Yang,
  Phys.\ Rev.\ C {\bf 51}, 1465 (1995).
  

\bibitem{Chung:1995dx} 
  S.~U.~Chung, J.~Brose, R.~Hackmann, E.~Klempt, S.~Spanier and C.~Strassburger,
  Annalen Phys.\  {\bf 4}, 404 (1995).

\bibitem{martin1970elementary}
   A.D. Martin, and T.D. Spearman, 
   {\it Elementary particle theory}
   (North-Holland Pub. Co., Amsterdam, 1970).

\bibitem{Lo:2017sde} 
  P.~M.~Lo,
  Eur.\ Phys.\ J.\ C {\bf 77}, 533 (2017).
 

\bibitem{Book_Byckling}
E.~Byckling and K. Kajantie, 
{\it Particle Kinematics}, 
(John Wiley \& Sons Ltd, 1973).
 
\bibitem{Sakurai:2011zz} 
  J.~J.~Sakurai and J.~Napolitano,
  {\it Modern quantum physics},
  (Boston, USA: Addison-Wesley, 2011).
  
\bibitem{Weinberg:1971mx} 
  S.~Weinberg,
  Astrophys.\ J.\  {\bf 168}, 175 (1971).

  
\bibitem{Patrignani:2016xqp} 
  C.~Patrignani {\it et al.} [Particle Data Group],
  Chin.\ Phys.\ C {\bf 40}, 100001 (2016).
\end{thebibliography}


\end{document}